\documentclass{aa}

\usepackage{graphicx}
\usepackage{txfonts}
\usepackage{natbib}
\usepackage{xcolor}
\usepackage{xspace}
\usepackage[colorlinks=true,linkcolor=blue,citecolor=blue]{hyperref}
\usepackage{etoolbox}

\makeatletter

\patchcmd{\NAT@citex}
  {\@citea\NAT@hyper@{%
     \NAT@nmfmt{\NAT@nm}%
     \hyper@natlinkbreak{\NAT@aysep\NAT@spacechar}{\@citeb\@extra@b@citeb}%
     \NAT@date}}
  {\@citea\NAT@nmfmt{\NAT@nm}%
   \NAT@aysep\NAT@spacechar\NAT@hyper@{\NAT@date}}{}{}

\patchcmd{\NAT@citex}
  {\@citea\NAT@hyper@{%
     \NAT@nmfmt{\NAT@nm}%
     \hyper@natlinkbreak{\NAT@spacechar\NAT@@open\if*#1*\else#1\NAT@spacechar\fi}%
       {\@citeb\@extra@b@citeb}%
     \NAT@date}}
  {\@citea\NAT@nmfmt{\NAT@nm}%
   \NAT@spacechar\NAT@@open\if*#1*\else#1\NAT@spacechar\fi\NAT@hyper@{\NAT@date}}
  {}{}

\makeatother

\newcommand{\hlink}[1]{\url{http://#1}\xspace}

\newcommand{\rfig}[1]{Fig.~\ref{#1}}
\newcommand{\rfigs}[1]{Figs.~\ref{#1}}
\newcommand{\req}[1]{Eq.~\ref{#1}}

\newcommand{\rtab}[1]{Table \ref{#1}}

\newcommand{\rapp}[1]{Appendix \ref{#1}}

\newcommand{\Rsec}[1]{Section \ref{#1}}
\newcommand{\rsec}[1]{section \ref{#1}}

\newcommand{\herschel}{{\it Herschel}\xspace}
\newcommand{\spitzer}{{\it Spitzer}\xspace}
\newcommand{\hubble}{{\it Hubble}\xspace}
\newcommand{\hst}{{\it HST}\xspace}
\newcommand{\jwst}{{\it JWST}\xspace}

\newcommand{\um}{\mu{\rm m}}

\newcommand{\uJy}{\mu{\rm Jy}}
\newcommand{\Jykms}{{\rm Jy.km/s}}
\newcommand{\kms}{{\rm km/s}}
\newcommand{\mJy}{{\rm mJy}}

\newcommand{\sfr}{{\rm SFR}}

\newcommand{\sfrms}{{\rm SFR}_{\rm MS}}

\newcommand{\lir}{L_{\rm IR}}
\newcommand{\lfir}{L_{\rm FIR}}
\newcommand{\lbol}{L_{\rm bol}}
\newcommand{\lion}{L_{E > 6{\rm eV}}}
\newcommand{\leight}{L_8}

\newcommand{\cplus}{{\rm [\ion{C}{II}]}\xspace}
\newcommand{\lcplus}{L_{[\ion{C}{ii}]}}
\newcommand{\fcplus}{S_{[\ion{C}{ii}]}}

\newcommand{\lsun}{L_\odot}
\newcommand{\msun}{{\rm M}_\odot}
\newcommand{\mgas}{M_{\rm gas}}
\newcommand{\mdyn}{M_{\rm dyn}}
\newcommand{\fgas}{f_{\rm gas}}
\newcommand{\Mpc}{{\rm Mpc}}
\newcommand{\kpc}{{\rm kpc}}
\newcommand{\pc}{{\rm pc}}
\newcommand{\Gyr}{{\rm Gyr}}
\newcommand{\Myr}{{\rm Myr}}
\newcommand{\yr}{{\rm yr}}
\newcommand{\dex}{{\rm dex}}
\newcommand{\mstar}{M_\ast}

\newcommand{\tdust}{T_{\rm dust}}

\newcommand{\galfit}{GALFIT\xspace}

\newcommand{\mdust}{M_{\rm dust}}
\newcommand{\kelvin}{{\rm K}}
\newcommand{\hyde}{{\it Hyde}\xspace}
\newcommand{\jekyll}{{\it Jekyll}\xspace}

\newcommand{\Ks}{$K_{\rm s}$\xspace}

\defcitealias{schreiber2015}{S15}
\defcitealias{chary2001}{CE01}
\defcitealias{dale2002}{DH02}
\defcitealias{galliano2011}{G11}
\defcitealias{draine2007}{DL07}

\newcommand*\dd{\ensuremath{d}}

\begin{document}

\title{Jekyll \& Hyde: quiescence and extreme obscuration \\ in a pair of massive galaxies 1.5 Gyr after the Big Bang}
\titlerunning{Jekyll \& Hyde: quiescence and extreme obscuration 1.5 Gyr after the Big Bang}

\author{C.~Schreiber\inst{1}
  \and I.~Labb\'e\inst{1}
  \and K.~Glazebrook\inst{2}
  \and G.~Bekiaris\inst{2,3}
  \and C.~Papovich\inst{4}
  \and T.~Costa\inst{1}
  \and D.~Elbaz\inst{5}
  \and G.~G.~Kacprzak\inst{2}
  \and T.~Nanayakkara\inst{1}
  \and P.~Oesch\inst{6}
  \and M.~Pannella\inst{7}
  \and L.~Spitler\inst{8,9,10}
  \and C.~Straatman\inst{11}
  \and K.-V.~Tran\inst{3,12}
  \and T.~Wang\inst{13,14}
}

\institute{
    Leiden Observatory, Leiden University, NL-2300 RA Leiden, The Netherlands \\
    \email{cschreib@strw.leidenuniv.nl} 
    \and Centre for Astrophysics and Supercomputing, Swinburne University of Technology, Hawthorn, VIC 3122, Australia 
    \and Australia Telescope National Facility, CSIRO Astronomy and Space Science, PO Box 76, Epping, NSW 1710, Australia 
    \and George P.~and Cynthia W.~Mitchell Institute for Fundamental Physics and Astronomy, Department of Physics and Astronomy, Texas A\&M University, College Station, TX 77843, USA 
    \and AIM-Paris-Saclay, CEA/DSM/Irfu -- CNRS -- Universit\'e Paris Diderot, CEA-Saclay, pt courrier 131, 91191 Gif-sur-Yvette, France 
    \and Observatoire de Gen\`eve, 1290 Versoix, Switzerland 
    \and Faculty of Physics, Ludwig-Maximilians Universit\"at, Scheinerstr.~1, 81679 Munich, Germany 
    \and Research Centre for Astronomy, Astrophysics \& Astrophotonics, Macquarie University, Sydney, NSW 2109, Australia 
    \and Department of Physics \& Astronomy, Macquarie University, Sydney, NSW 2109, Australia 
    \and Australian Astronomical Observatory, 105 Delhi Rd., Sydney NSW 2113, Australia 
    \and Max-Planck Institut f\"ur Astronomie, K\"onigstuhl 17, D-69117, Heidelberg, Germany 
    \and School of Physics, University of New South Wales, Sydney, NSW 2052, Australia 
    \and Institute of Astronomy, The University of Tokyo, Osawa, Mitaka, Tokyo 181-0015, Japan 
    \and National Astronomical Observatory of Japan, Mitaka, Tokyo 181-8588, Japan 
}

\date{Received 08 September 2017; accepted 07 December 2017}
{
\abstract {We obtained ALMA spectroscopy and deep imaging to investigate the origin of the unexpected sub-millimeter emission toward the most distant quiescent galaxy known to date, ZF-COSMOS-20115 at $z=3.717$. We show here that this sub-millimeter emission is produced by another massive ($\mstar \sim 10^{11}\,\msun$), compact ($r_{1/2}=0.67\pm0.14\,\kpc$) and extremely obscured galaxy ($A_{\rm V} \sim 3.5$), located only $0.43\arcsec$ ($3.1\,\kpc$) away from the quiescent galaxy. We dub the quiescent and dusty galaxies \jekyll and \hyde, respectively. No dust emission is detected at the location of the quiescent galaxy, implying $\sfr < 13\,\msun/\yr$ which is the most stringent upper limit ever obtained for a quiescent galaxy at these redshifts. The two sources are spectroscopically confirmed to lie at the same redshift thanks to the detection of $\cplus_{158}$ in \hyde ($z=3.709$), which provides one the few robust redshifts for a highly-obscured ``$H$-dropout'' galaxy ($H-[4.5] = 5.1 \pm 0.8$). The \cplus line shows a clear rotating-disk velocity profile which is blueshifted compared to the Balmer lines of \jekyll by $549\pm60\,\kms$, demonstrating that it is produced by another galaxy. Careful de-blending of the \spitzer imaging confirms the existence of this new massive galaxy, and its non-detection in the \hubble images requires extremely red colors and strong attenuation by dust. Full modeling of the UV-to-far-IR emission of both galaxies shows that \jekyll has fully quenched at least $200\,\Myr$ prior to observation and still presents a challenge for models, while \hyde only harbors moderate star-formation with $\sfr \lesssim 120\,\msun/\yr$, and is located at least a factor $1.4$ below the $z\sim4$ main sequence. \hyde could also have stopped forming stars less than $200\,\Myr$ before being observed; this interpretation is also suggested by its compactness comparable to that of $z\sim4$ quiescent galaxies and its low \cplus/FIR ratio, but significant on-going star-formation cannot be ruled out. Lastly, we find that despite its moderate $\sfr$, \hyde hosts a dense reservoir of gas comparable to that of the most extreme starbursts. This suggests that whatever mechanism has stopped or reduced its star-formation must have done so without expelling the gas outside of the galaxy. Because of their surprisingly similar mass, compactness, environment and star-formation history, we argue that \jekyll and \hyde can be seen as two stages of the same quenching process, and provide a unique laboratory to study this poorly understood phenomenon.
}


\maketitle

\section{Introduction}

In the local Universe, more than half of the stellar mass is found in quiescent galaxies \citep[e.g.,][]{bell2003} with current star-formation rates ($\sfr$s) only $\lesssim$1\% of their past average (e.g., \citealt{pasquali2006}). Unlike star-forming galaxies, which are predominantly rotating disks, quiescent galaxies have spheroidal shapes, very dense stellar cores, and dispersion-dominated kinematics. They contain very little atomic and molecular gas (e.g., \citealt{combes2007,saintonge2016}), and most of their gas is instead ionized (e.g., \citealt{annibali2010}). These galaxies also frequently possess an active galactic nucleus (AGN; e.g., \citealt{lee2010}), signpost of accretion of matter onto a central super-massive black hole. Lastly, they tend to be much rarer at low stellar masses \citep[e.g.,][]{baldry2004}, and more abundant in dense environments \citep[e.g.,][]{peng2010}.

The formation channel of such galaxies remains uncertain. A number of mechanisms have been proposed to stop (i.e., ``quench'') or reduce star-formation, and all effectively act to deplete the cold gas reservoirs. This can achieved by a) removing cold gas from the galaxy through outflows, b) pressurizing the gas and preventing it from collapsing, c) stopping the supply of infalling gas until star-formation exhausts the available reserves, or d) any combination of these. The underlying physical processes could be various, including feedback from stars or an AGN, injection of kinetic energy from infalling gas, stabilization of a gas disk by a dense stellar core, or tidal interactions with massive neighboring galaxies (e.g., \citealt{silk1998,birnboim2003,croton2006,gabor2012,martig2009,foersterschreiber2014,genzel2014,peng2015}). While there is evidence that each of these phenomena does (or can) happen in at least some galaxies, it still remains to be determined which of them actually plays a significant role in producing the observed population of quiescent galaxies.

At higher redshifts, where spectroscopy is scarce and more expensive, selecting quiescent galaxies is challenging. Yet this has proven to be a powerful tool to constrain their formation mechanism and the overall process driving the growth of galaxies in general \citep[e.g.,][]{peng2010}. In the absence of spectroscopy, selection criteria based on red broad-band colors have been designed, preferably with two colors to break the age-dust degeneracy \citep{franx2003,labbe2007,williams2009,arnouts2013}. Using such methods, it was found that quiescent galaxies were less numerous in the past \citep[e.g.,][]{labbe2007,muzzin2013,ilbert2013,tomczak2014}, consistent with the fact that this population has been slowly building up with time. Surprisingly, quiescent and massive galaxies are still found up to very high redshifts \citep{glazebrook2004,straatman2014}, implying that star-formation may be more rapid and quenching more efficient than envisioned by most models.

Yet, spectroscopic confirmation of their redshifts and quiescence is required to draw firm conclusions \citep[e.g.,][]{kriek2009,gobat2012}. In \cite{glazebrook2017} we reported the spectroscopic identification of the most distant quiescent galaxy known to date, ZF-COSMOS-20115, at $z=3.717$. The galaxy was first identified in \cite{straatman2014} thanks to its strong Balmer break, and its redshift was subsequently confirmed using deep Balmer absorption lines, a clear indicator of a recent shutdown of star-formation. This allowed us to precisely trace back the star-formation history of this galaxy, which we estimated must have stopped at $z>5$ and required a large peak $\sfr\sim1000\,\msun/\yr$. While some models can accommodate quenched galaxies this early in the history of the Universe \citep[e.g.,][]{rong2017,qin2017}, none is able to produce them in numbers large enough to match observations.

Despite its apparently quenched star-formation history, faint sub-millimeter emission was detected toward the galaxy ZF-COSMOS-20115 with ALMA \citep{schreiber2017-b}, with a spatial offset of $0.4\pm0.1\arcsec$. This suggested that star-formation might still be on-going in an obscured region of the galaxy, and would thus have escaped detection at shorter wavelengths. As discussed in \cite{glazebrook2017}, this emission is faint enough that the corresponding $\sfr$, even if indeed associated with the quiescent galaxy, would only account for up to $15\%$ of the total observed stellar mass over the last $200\,\Myr$. Therefore, regardless of the sub-millimeter emission, the bulk of the mass in this galaxy had necessarily formed earlier on. But the questions remained of whether the galaxy has actually quenched, and what is truly powering the sub-millimeter emission (see also \citealt{simpson2017}).

To answer these questions, we have obtained deeper and higher-resolution continuum imaging at $744\,\um$ with ALMA, in a spectral window centered on the expected frequency of $\cplus_{158}$ at $z=3.717$. The present paper discusses these new observations and what they reveal about the true nature of this system. In the following, we refer to the quiescent galaxy as ``\jekyll'' and the sub-millimeter source as ``\hyde''.

The flow of the paper goes as follows. In \rsec{SEC:obs:alma} we describe the dust and \cplus data we used in this paper, and how we modeled them. The relative astrometry between ALMA and \hubble is quantified in \rsec{SEC:astrometry}, and the flux extraction is detailed in \rsec{SEC:flux_extract}. The dust emission modeling and results are described in \rsec{SEC:dustprops}, while the modeling of the kinematics of the \cplus line is addressed in \rsec{SEC:disk_model}. We then study the stellar emission of \jekyll and \hyde in \rsec{SEC:obs:uvnir}, starting from the UV to near-IR photometry in \rsec{SEC:photometry}, following by a description of the spectral modeling in \rsec{SEC:stellar_fit} and a description of the results in \rsec{SEC:stellar_pop}. In \rsec{SEC:discussion} we put together and discuss these observations. \Rsec{SEC:dusty_psb} addresses the non-detection of the dust continuum in \jekyll and its quiescent nature, and \rsec{SEC:hyde_galaxy} demonstrates that \hyde is indeed a separate galaxy. We then study evidence for a recent or imminent quenching in \hyde, based on its compactness in \rsec{SEC:size} and low \cplus luminosity in \rsec{SEC:cplus_deficit}. The efficiency of star-formation and possible quenching processes are discussed in \rsec{SEC:sfe}. Lastly, \rsec{SEC:past_jekyll} speculates on how \hyde can be viewed as a younger version of \jekyll, and what links can be drawn between the two galaxies. We then summarize our conclusions in \rsec{SEC:conclusion}.

In the following, we assumed a $\Lambda$CDM cosmology with $H_0 = 70\ {\rm km}\,{\rm s}^{-1} {\rm Mpc}^{-1}$, $\Omega_{\rm M} = 0.3$, $\Omega_\Lambda = 0.7$ and a \cite{chabrier2003} initial mass function (IMF), to derive both star-formation rates and stellar masses. All magnitudes are quoted in the AB system, such that $M_{\rm AB} = 23.9 - 2.5\log_{10}(S_{\!\nu}\ [\uJy])$.

\section{Dust continuum and \cplus emission \label{SEC:obs:alma}}

The first observations toward this system with ALMA were obtained at $338\,{\rm GHz}$ ($888\,{\rm \mu m}$) in band 7, targeting \jekyll as part of a larger program (2013.1.01292.S, PI: Leiton) observing massive $z\sim4$ galaxies \citep{schreiber2017-b}. The on-source observing time was only $1.5$ minutes, and the resulting noise level reached $0.25\,\mJy$ ($1\sigma$) with a beam of full width at half maximum of $1.1\times0.7\arcsec$. Nevertheless, a source was detected at $5\sigma$ significance with a flux of $1.52\pm0.25\,\mJy$, slightly offset from the position of \jekyll ($0.5\pm0.1\arcsec)$. This detection was already discussed in our previous work where we introduced \jekyll \citep{glazebrook2017}. At the time, the spatial offset was already deemed significant, although the limited signal-to-noise ratio of the ALMA source as well as the relatively wide beam were such that its interpretation was difficult.

Once the ALMA emission was discovered, we proposed to re-observe this system with better sensitivities and angular resolution, and sought to detect the \cplus line to confirm the redshift. These observations are described in the next sections, together with the flux extraction and modeling procedure (both for the ALMA data and ancillary imaging from \spitzer and \herschel). All our results are summarized in \rtab{TAB:props}.

\begin{figure*}
\begin{center}
\includegraphics[width=\textwidth]{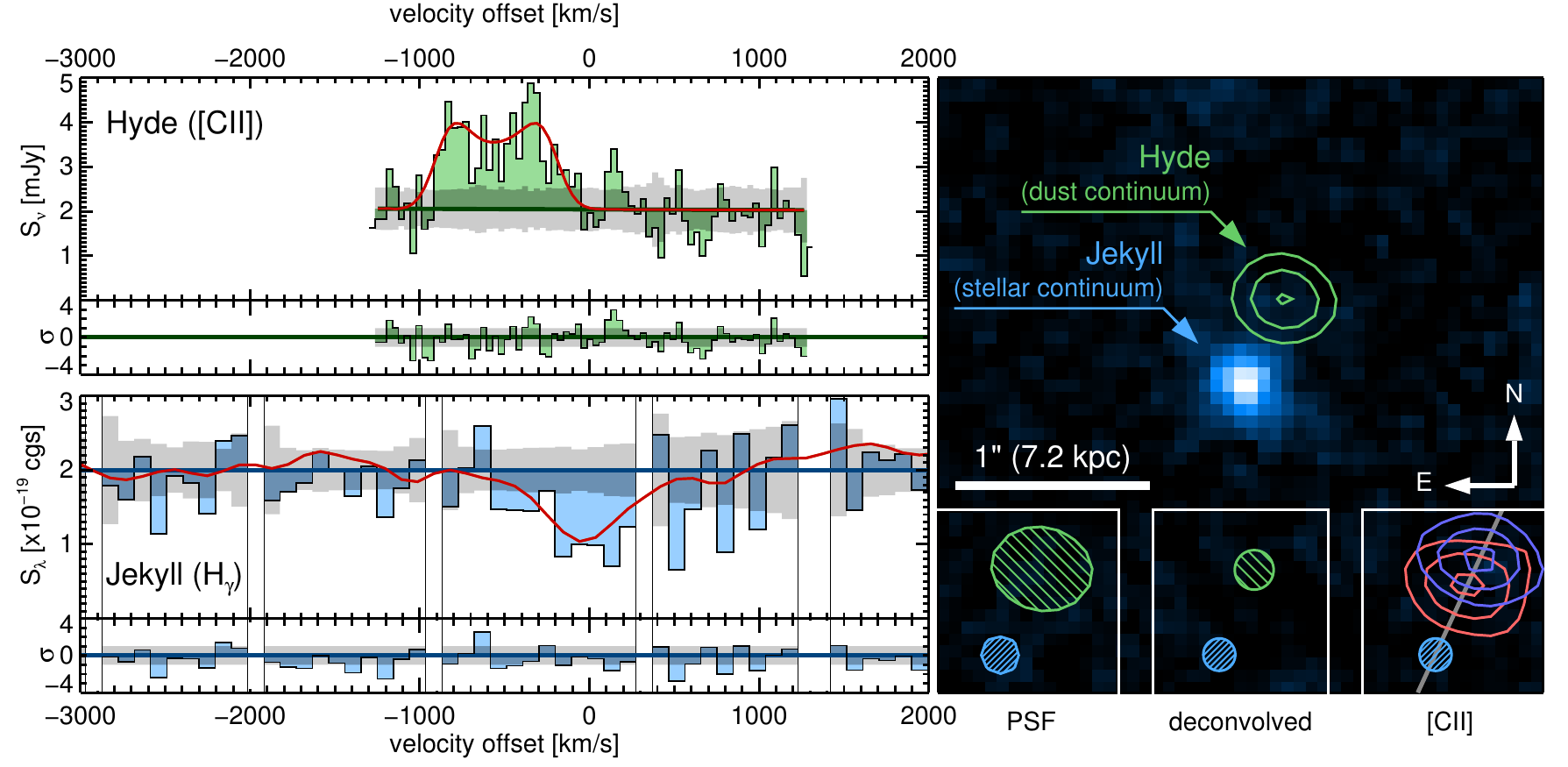}
\end{center}
\caption{Spectra and imaging of \jekyll \& \hyde. {\bf Top left:} ALMA spectrum of \hyde. {\bf Bottom left:} MOSFIRE spectrum of \jekyll (binned to $6.5\,\AA$ resolution). The two spectra are shown on the same velocity scale. The emission above and below the continuum level is shaded to emphasize the lines. The gray shaded area in the background is the 1$\sigma$ flux uncertainty. We show the models best fitting these spectra with red lines. At the bottom of each plot we give the normalized model residual $\sigma$, i.e., the difference between observed and modeled flux divided by the uncertainty. {\bf Right:} image of the \jekyll \& \hyde system. The background image (blue tones; false colors) is the near-IR $H$-band emission as observed by \hubble; the bright source at the center is \jekyll. We overlay the ALMA $740\,\um$ continuum emission with green contours (the most extended contour corresponds half of the peak emission); the source detected here is \hyde. The full width at half maximum of the {\it Hubble} and ALMA point spread functions are given on the bottom left corner, followed to the right by the deconvolved profiles of the two galaxies (half-light area). Lastly, the inset on the bottom right corner shows the position of the two velocity components of \hyde with respect to \jekyll (the blue contours correspond to the most blueshifted component), and a gray line connects the two galaxies. \label{FIG:cutout}}
\end{figure*}

\subsection{Overview of ALMA observations and flux extraction}


The new ALMA data were obtained in band 8 with a single spectral tuning at $403\,{\rm GHz}$ ($744\,{\rm \mu m}$), and were delivered in early 2017 as part of the Director's Discretionary Time (DDT) program 2015.A.00026.S. The on-source observing time was $1.2$ hours, with a beam size of $0.52\times0.42\arcsec$ (natural weighting), about a factor of two smaller than the first observations. The pointing was centered on the {\it Hubble Space Telescope} (\hst) position of \jekyll: $\alpha=150.06146\degr$, $\delta=2.378683\degr$.

We generated two spectral cubes in CASA corresponding to two pairs of spectral windows of disjoint frequency range, the first centered on the expected frequency of the \cplus line ($401.2$ to $404.7\,{\rm GHz}$), and the second at higher frequencies which only measure the continuum level ($413.0$ to $416.5\,{\rm GHz}$). We did not perform any cleaning on these cubes, and thus used the dirty images throughout this analysis with the dirty beam as point-spread function. We binned the flux of every three frequency channels to eliminate correlated noise between nearby channels, and determined the noise level in each channel using the RMS of the pixels away from the source (without applying the primary-beam correction). We found that the frequencies $402.4$ and $414.3\,{\rm GHz}$ are affected by known atmospheric lines which increase the noise by a factor of two, the former can be seen on \rfig{FIG:cutout} (top-left) at $+400\,\kms$.

To extract the continuum and line flux of the target, we proceeded as follows. We first created a ``continuum+line'' image by averaging all spectral channels together, and located the peak position of the emission. A bright source was found, with a peak flux of $2.36\pm0.06\,\mJy$, and clearly offset from the position of the quiescent galaxy by about $0.5\arcsec$. The accuracy of the astrometry is demonstrated in \rsec{SEC:astrometry}, and the offset is discussed further in \rsec{SEC:flux_extract}. We then extracted a spectrum, shown in \rfig{FIG:cutout} (top-left), at this peak position and found a line close to the expected frequency of \cplus at $z\sim3.7$ (throughout this paper, we assumed a vacuum rest frequency $\nu_{\rm \cplus\,rest} = 1900.5369\,{\rm GHz}$, which known with an excellent accuracy of $\Delta \nu = 1.3\,{\rm MHz}$; \citealt{cooksy1986}). The line is relatively broad and its kinematics resembles more a ``double horn'', typical of rotating disks, than a single Gaussian.

We created a continuum image by masking the spectral channels containing the \cplus emission. From this image we extracted the size and total continuum flux of the source, as described in \rsec{SEC:flux_extract}. This flux and the ancillary \herschel and SCUBA-2 photometry is modeled in \rsec{SEC:dustprops}. We then subtracted the continuum map from the spectral cube and fit the \cplus emission with a rotating disk model described in \rsec{SEC:disk_model}. This model was used to determine the spatial extent, total flux, and kinematics of the \cplus emission.

\subsection{ALMA astrometry \label{SEC:astrometry}}


Given the $S/N=40$ of the detection in the new ALMA image, the position of the dust emission is known with an uncertainty of $0.01\arcsec$. Since the \hubble imaging of \jekyll also provides a high $S/N$ detection and shows that \jekyll is very compact ($r_{1/2} = 0.07\pm0.02\arcsec$; \citealt{straatman2015}), the two sources are undoubtedly offset. However, since \jekyll and \hyde are each detected by a different instrument, it is possible that either image is affected by a systematic astrometric issue which could spuriously generate such an offset. For example \cite{rujopakarn2016} and \cite{dunlop2017} have revealed that the \hubble imaging in the GOODS--South field was affected by a systematic astrometry shift of about $0.26\arcsec$, when compared to images from ALMA, VLA and 2MASS. The same study also showed that VLA and ALMA positions match within $0.04\arcsec$.

The two galaxies discussed in the present paper are located in the COSMOS field. Here, all the UV-to-NIR images (including that from \hubble) are tied to a common astrometric frame defined by the CFHT $i^*$-band image, which itself is ultimately anchored to the radio image from the VLA \citep{koekemoer2007}. Since both ALMA and VLA have been shown to provide consistent absolute astrometry, we do not expect such large offsets to exist in the COSMOS field. A comparison of the \hubble astrometry against Pan-STARRS suggests indeed that no large offset exists in this field (M.~Dickinson, private communication).

To confirm this, we have retrieved from the ALMA archive\footnote{Projects 2013.1.00034.S, 2013.1.00118.S, 2013.1.00151.S, 2015.1.00137.S, 2015.1.00379.S and 2015.1.01074.S.} the images of sub-millimeter galaxies in the COSMOS field, observed in bands 6 and 7, and which have a clear counterpart in the VISTA \Ks image within a $1.5\arcsec$ radius. We chose the \Ks-band image as a reference instead of the \hst $H$ band since it provides the best $S/N$ for this sub-millimeter sample, and also because it covers the entire COSMOS field. Since both the \Ks and \hst images are tied to the same astrometric reference (see above), the results we discuss below also apply to the \hst imaging.

For each source, we estimated the uncertainty in the ALMA and VISTA centroids, which depend on the signal-to-noise ratio ($S/N$) of the source and the size of the beam or point spread function ($\sigma_{\rm beam}$) as $\Delta p_{N, \rm X} = \sqrt{2}\,\sigma_{\rm beam}/(S/N)$. Defining $\Delta p_N^2 = \Delta p_{N, \rm ALMA}^2 + \Delta p_{N, \rm VISTA}^2$, we then excluded sources with $\Delta p_N>0.15\arcsec$ to only consider the well-measured centroids. To avoid blending issues, we excluded from this sample the sources detected as more than one component in the \Ks image or with a neighbor within $2\arcsec$. To avoid physical offsets caused by patchy obscuration, we also excluded galaxies with a detection on the Subaru $i$ image significantly offset from the \Ks centroid. This adds up to a sample of $71$ galaxies. For each source, we measured the position of both the ALMA and \Ks sources as the flux weighted average of the $x$ and $y$ coordinates and computed the positional offset between the two.

The resulting absolute offsets are displayed in \rfig{FIG:alma_dpos}. We found an average of $\Delta\alpha=+0.068\pm0.012\arcsec$ and $\Delta\delta=-0.031\pm0.013\arcsec$ ($\rm ALMA - VISTA$), which is confined to less than $0.1\arcsec$ but nevertheless significant (see also \citealt{smoli2017} where a similar offset was reported). However, these numbers only apply to the COSMOS field as a whole (our sample spans $1.2\degr\times1.1\degr$); systematic offsets may vary spatially, but average out when computed over the entire field area. To explore this possibility, we selected only the galaxies that lie within $5\arcmin$ of our objects, reducing the sample to six galaxies. In this smaller but more local sample, we found averages of $\Delta\alpha=+0.11\pm0.03\arcsec$ and $\Delta\delta=+0.04\pm0.04\arcsec$, which are consistent with the previous values thus imply no significant variation across the field. For all the following, we therefore assumed the global offset derived above and brought the ALMA positions back into the same astrometric reference as the optical-NIR images.

After subtracting this small systematic offset, the largest offset we observed in the full sample of $71$ galaxies was $0.33\arcsec$, and $0.20\arcsec$ in the smaller sample of six galaxies. These values are both lower than the $\sim0.5\arcsec$ offset we observed between \jekyll and \hyde and suggest that the latter being offset by chance is unlikely. We quantify this in the next paragraph.

\begin{figure}
\begin{center}
\includegraphics[width=0.49\textwidth]{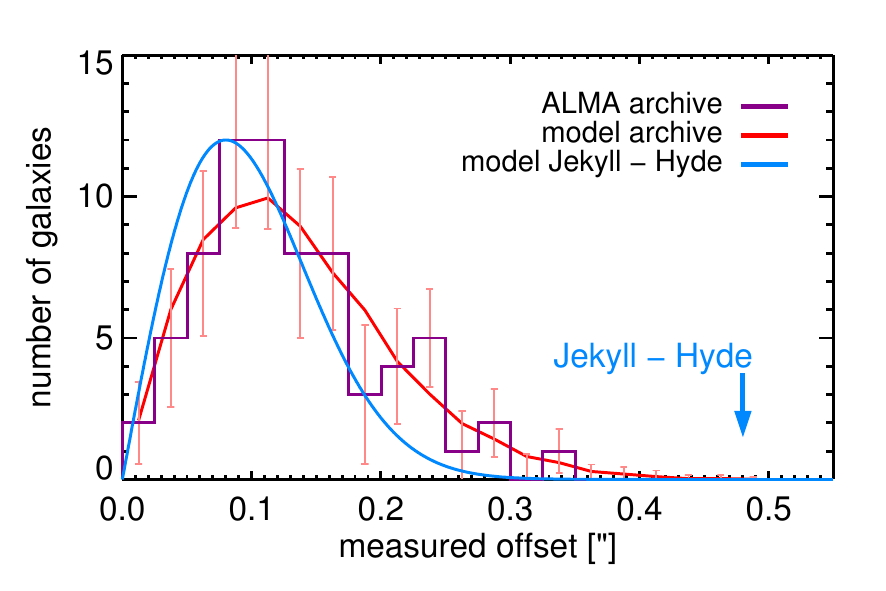}
\end{center}
\caption{Distribution of observed positional offsets between ALMA and VISTA in the COSMOS field. The purple histogram shows the observed distribution for $76$ galaxies selected from the ALMA archive, and the red line is the best-fit model (including telescope pointing accuracy and uncertainty in the centroid determination on noisy images). Error bars show counting uncertainties derived assuming Poisson statistics from the best-fit model. The blue arrow shows the offset observed between \jekyll and \hyde, and the blue line is the expected offset distribution given the $S/N$ and PSF width of the two galaxies on their respective images. \label{FIG:alma_dpos}}
\end{figure}

To determine the random astrometric registration errors between ALMA and VISTA, we modeled the observed offsets using two sources of offsets (per coordinate). On the one hand, we considered random offsets caused by noise in the ALMA and VISTA images, the amplitude of which are given by $\Delta p_N$ as described above. On the other hand, we considered the combined pointing accuracy of ALMA and VISTA $\Delta p_T$, which we assumed is a constant value identical for both coordinates. The total uncertainty on a coordinate of the source $i$ is then $\Delta p(i)^2 = \Delta p_N(i)^2 + \Delta p_T^2$. Varying $\Delta p_T$ on a grid from $0$ to $0.5\arcsec$, we generated $200$ Monte Carlo simulations of the sample and compared the simulated offset distribution to the observed one using a Kolmogorov-Smirnov test. We found $\Delta p_T = 0.080\pm0.009\arcsec$, and display the best-fit model in \rfig{FIG:alma_dpos}. The probability of observing a given offset by chance is then governed by a Rayleigh distribution of scale parameter $\Delta p(i)$. In the case of \jekyll and \hyde, the $S/N$ in both VISTA and ALMA is high such that $\Delta p_N = 0.02\arcsec \ll \Delta p_T$. This implies the probability of observing an offset of $\ge0.4\arcsec$ by chance is only $10^{-4}$ (see \rfig{FIG:alma_dpos}), and even less if we consider that the offset is observed independently in both band 7 and band 8 images. The observed offset is therefore real.

\subsection{Fluxes and spatial profiles \label{SEC:flux_extract}}


\begin{figure}
\begin{center}
\includegraphics[width=0.49\textwidth]{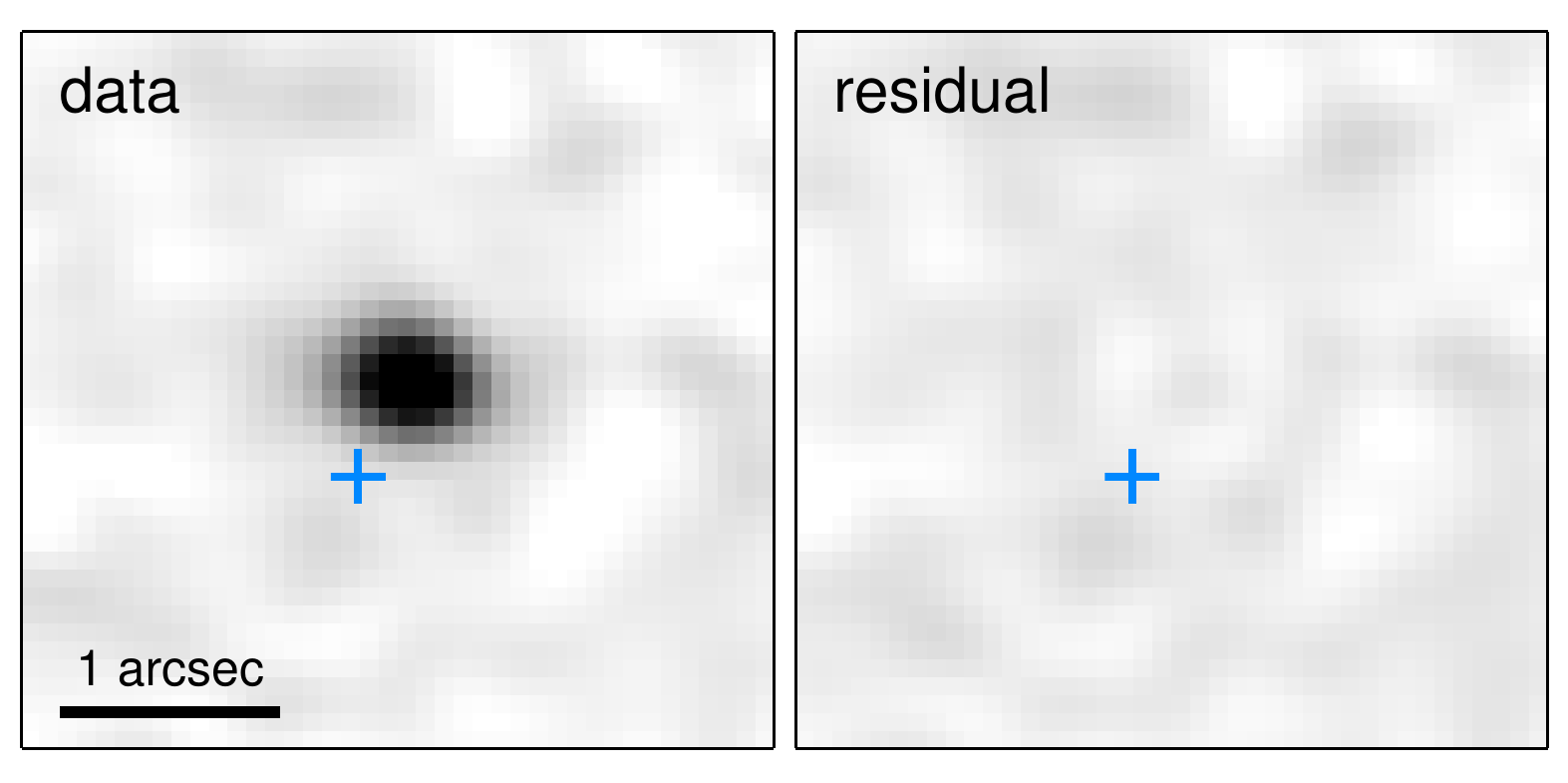}
\end{center}
\caption{ALMA $744\,\um$ continuum emission (left) and residual (right) after subtracting the best-fit exponential disk model with \texttt{imfit}. The centroid of the \hst emission of \jekyll is indicated with a blue cross. The beam FWHM is $0.52\times0.42\arcsec$. \label{FIG:alma_resid}}
\end{figure}

We used \texttt{imfit}\footnote{\url{https://github.com/perwin/imfit}} v1.5 \citep{erwin2015} to model the dust continuum emission, assuming an exponential disk profile (\citealt{hodge2016}) and Gaussian noise. Since we model the dirty image directly, the correct point-spread function to use in the modeling is the dirty beam. However since this beam has a zero integral, one should not re-normalize it at any stage of the modeling. We therefore had to disable the re-normalization of the PSF in \texttt{imfit} using the \texttt{-{}-no-normalize} flag. We cross-checked our results by modeling the continuum emission in line-free channels using both \texttt{uvmodelfit} and \texttt{uvmultifit} \citep{marti-vidal2014}, which both analyze the emission directly in the $(u,v)$ plane rather than on reconstructed images, and found similar results.

To compute uncertainties in the model parameters, we ran \texttt{imfit} on simulated data sets with the same noise as the observed data (i.e., a white Gaussian noise map convolved with the dirty beam and re-normalized to the RMS of the observed image), where we artificially injected a source with the same size and flux as our best-fit model. The uncertainties were then determined from the standard deviation of the best fits among all simulated data sets.

The ALMA emission and residual are shown in \rfig{FIG:alma_resid}. We measured for \hyde a total continuum flux of $S_{744\,\um}=2.31\pm0.14\,\mJy$, offset from \jekyll by $\Delta\alpha = -0.132\pm0.017\arcsec$ and $\Delta\delta = +0.405\pm0.015\arcsec$, which is consistent with the offset previously measured in the shallower data. This corresponds to a projected distance of $0.426\pm0.015\arcsec$, i.e., $3.05\pm0.11\,\kpc$. We showed in \rsec{SEC:astrometry} that this offset is highly significant: the dust emission must therefore originate from another object, \hyde. This source is marginally resolved, with a half-light radius of $0.10\pm 0.02\arcsec$ (i.e., the source is about half the size of the dirty beam). At $z=3.7$, this corresponds to $0.67\pm0.14\,\kpc$.

No significant continuum emission is detected at the location of \jekyll ($0.09\pm0.06\,\mJy$, assuming a point source, and accounting for de-blending and astrometry uncertainty using the procedure described in \rsec{SEC:photometry}). As illustrated in \rfig{FIG:cutout} (right), the projected distance between \jekyll and \hyde is much larger than their respective half-light radii (by a factor $\sim$$5$), therefore the two galaxies do not overlap and form two separate systems.

The far-IR photometry toward the \jekyll+\hyde system was re-extracted from \spitzer MIPS and \herschel imaging following a method standard to deep fields \citep{elbaz2011}, and briefly summarized below. Given the large beam sizes, it is impossible to de-blend \jekyll and \hyde on these images. Motivated by the fact that \jekyll is at least $\sim$$20$ times fainter than \hyde on the $744\,\um$ image, we assumed that the entirety of the MIPS and \herschel fluxes is produced by \hyde.

To account for the poor angular resolution of far-IR images, we modeled all sources in the ancillary images within a $5\arcmin\times5\arcmin$ region centered on the system. The $24$ to $160\,\um$ images were modeled with point-like sources at the position of \spitzer IRAC-detected galaxies. The $250$ to $500\,\um$ images were modeled similarly, using positions of \spitzer MIPS-detected galaxies. However since this provided a too high density of prior positions, we performed a second pass where MIPS priors with $250\,\um$ flux less than $3\,\mJy$ or negative $500\,\um$ were discarded. \hyde was always kept in the prior list. The SCUBA2 $450\,\um$ flux was taken from \cite{simpson2017} assuming no significant contamination by neighboring sources.

\subsection{Far-IR photometry and modeling \label{SEC:dustprops}}


\begin{figure}
\begin{center}
\includegraphics[width=0.5\textwidth]{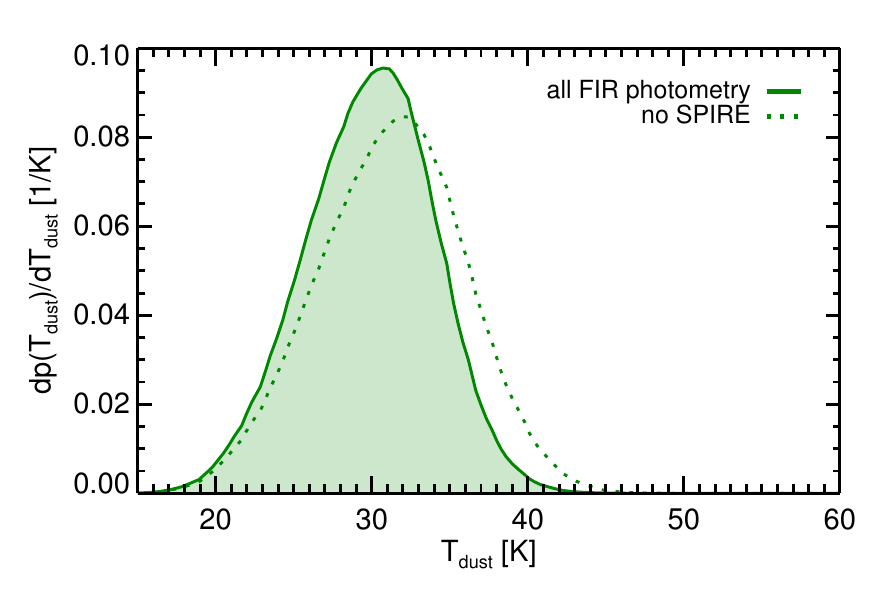}
\end{center}
\caption{Probability distribution of the dust temperature ($\tdust$) for \hyde. This was derived from the $\chi^2$ of a grid of $\tdust$ values tested against the observed photometry. The solid line shows the distribution using all the FIR photometry, and the dotted line shows how the distribution would have changed if we had not used the \herschel SPIRE photometry. \label{FIG:tdust_chi2}}
\end{figure}

We modeled the $24$--$890\,\mu{\rm m}$ photometry with the simple dust model presented in \cite{schreiber2017-a}, and here we briefly recall its main features. This model has three varying parameters: the dust temperature ($T_{\rm dust}$), the infrared luminosity ($L_{\rm IR}$, integrated from $8$ to $1000\,\mu{\rm m}$) and the $8\,\um$ luminosity ($\leight$). These templates are designed to describe the far-IR SED of star-forming galaxies with the best possible accuracy given this small number of free parameters. They are built from first principles using the dust model of \cite{galliano2011}, and therefore a dust mass ($\mdust$) is also associated to each template in the library. Compared to simpler gray-body models, these templates can accurately describe the emission at wavelengths shorter than the peak of the dust emission.

In the present case, since our data did not constrain the rest-frame $8\,\um$ emission, we fixed $\lir/\leight=8$, which is the value observed for massive star-forming galaxies at $z\sim2$ \citep{reddy2012-a,schreiber2017-a}. The fit therefore had four degrees of freedom. Before starting the fit, we subtracted from the observed $24\,\mu{\rm m}$ flux the estimated contribution from stellar continuum ($3.5\,\mu{\rm Jy}$), which we extrapolated from the best-fitting stellar continuum model (\rsec{SEC:stellar_fit}). Varying the dust temperature, we adjusted the infrared luminosity to best fit the observed data, and chose as best fit the dust temperature leading to the smallest reduced $\chi^2$. Uncertainties on all parameters were then computed by randomly perturbing the photometry within the error bars and re-doing the fit $5000$ times.

The resulting photometry is shown in \rfig{FIG:seds} (left) along with our best model. We found a dust temperature of $\tdust = 31^{+3}_{-4}\,\kelvin$ and a luminosity $\lir=1.1^{+0.4}_{-0.3}\times10^{12}\,\lsun$ (error bars include the uncertainty on $\tdust$) which is similar to that obtained by \cite{simpson2017}. This corresponds to $\sfr_{\rm IR} = 110^{+43}_{-33}\,\msun/\yr$ using the \cite{kennicutt1998-a} conversion, adapted to the Chabrier IMF following \cite{madau2014}. The dust mass is $\mdust = 3.2^{+2.2}_{-1.0}\times10^{8}\,\msun$, and is less well constrained than $\lir$ owing to the uncertainty on the dust temperature; the coverage of the dust SED at high and low frequency would need to be improved.

Similar values of $\lir=1.4\times10^{12}\,\lsun$ and $\sfr_{\rm IR}=140\,\msun/\yr$ were obtained by simply rescaling the SED of the $z=4.05$ starburst GN20 \citep{tan2014}, which has a similar dust temperature. In fact, significantly hotter temperatures were ruled out by the non-detections in all \herschel bands and the low SCUBA2 $450\,\um$ flux, see \rfig{FIG:tdust_chi2}. For example, assuming $\tdust=40\,\kelvin$ would have resulted in a combined $2.7\sigma$ tension with the observed photometry. Excluding the SPIRE fluxes, which are notoriously difficult to measure, we obtained a similar $\tdust = 32^{+4}_{-5}\,\kelvin$.

Given that \jekyll is not detected in any FIR image, we had to make an assumption on its dust temperature before interpreting its absence on the deep band 8 image. Rather than arbitrarily picking one temperature, we assumed a range of temperatures to obtain more conservative error bars. We considered $\tdust$ ranging from $20\,\kelvin$ (as observed in $z=2$ quiescent galaxies; \citealt{gobat2017}) to $35\,\kelvin$ (the upper limit for \hyde) with a uniform probability distribution. With this assumption, the non-detection of \jekyll in the band 8 image translates into $\lir = 3.6^{+3.1}_{-2.4}\times10^{10}\,\lsun$, or a $3\sigma$ upper limit of $\sfr_{\rm IR} < 13\,\msun/\yr$. This is the strongest upper limit ever obtained for a single quiescent galaxy at these redshifts \citep{straatman2014}, and is consistent with its quiescent nature derived from the SED modeling, which we revisit in \rsec{SEC:stellar_pop}. We note that even if we had assumed a high temperature of $\tdust = 40\,\kelvin$, which is substantially hotter than \hyde, the limits on $\lir$ and $\sfr_{\rm IR}$ would still be low: $\lir=(1.0\pm0.7)\times10^{11}\,\lsun$ and $\sfr_{\rm IR}<31\,\msun/\yr$ ($3\sigma$). Yet we consider such high temperatures unlikely; as we demonstrate later in \rsec{SEC:stellar_pop}, with only $A_{\rm V}=0.2$--$0.5\,{\rm mag}$ the large stellar mass of \jekyll is enough to reach $\lir\sim10^{11}\,\lsun$ without on-going star-formation (heating the dust with intermediate-age stars). This leaves little room for dust-obscured star-formation, in which case the dust must be cooler than typically observed in star-forming galaxies (e.g., \citealt{gobat2017}).

\subsection{Rotating disk model \label{SEC:disk_model}}

\begin{figure}
\begin{center}
\includegraphics[width=0.5\textwidth]{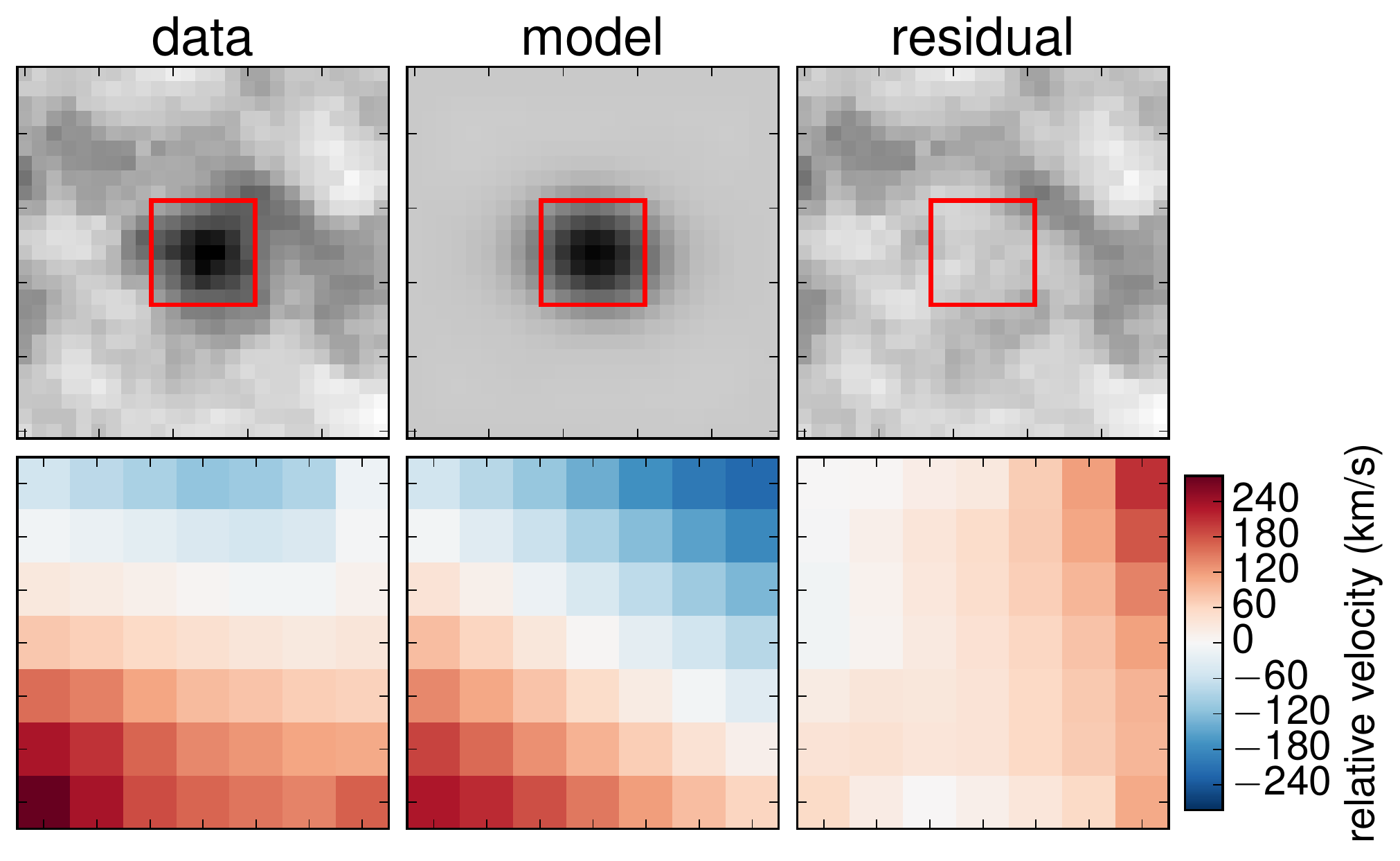}
\end{center}
\caption{Result of the modeling of the \cplus line emission with a rotating disk model. The first row shows the spectrally-integrated line intensity map, and the second row is the velocity field in the central region (indicated with a red box in the first row, $0.6\arcsec\times0.6\arcsec$). The beam FWHM is $0.52\times0.42\arcsec$. \label{FIG:disk_model_maps}}
\end{figure}

Since the velocity profile of the \cplus line shows a double-peaked structure, we modeled the continuum-subtracted spectral cube with an inclined thin disk model using \texttt{GBKFIT}\footnote{\url{https://github.com/bek0s/gbkfit}} \citep{bekiaris2016}. We fixed the centroid of the disk to that of the dust continuum, and varied the scale length ($h=10^{-5}$ to $3\,\kpc$), the inclination ($i=5$ to $85\degr$), the position angle (${\rm PA} = -90$ to $90\degr$), the central surface brightness ($I_0=0.02$ to $22\,\mJy/\kpc^2$), the systemic velocity ($v_{\rm sys} = 300$ to $700\,\kms$), the velocity dispersion ($\sigma_v=30$ to $200\,\kms$), and the turnover radius ($r_t = 10^{-4}$ to $6\,\kpc$) and velocity ($v_t=1$ to $1000\,\kms$).

For each combination of these parameters, we computed the total flux $S_{\cplus} = 2\pi\,I_0\,h^2\cos(i)$, the half-light radius $r_{\cplus} = 1.68\,h$, the velocity at $2.2\,h$, $v_{2.2} = (2\,v_t/\pi)\arctan(2.2\,h/r_t)$, the rotation period (or orbital time) $t_{\rm rot} = 2\pi\,2.2\,h/v_{2.2}$ and the dynamical mass $\mdyn = 2.2\,h\,{v_{2.2}}^2/G$.

The model best-fitting the observations was determined using a Levenberg-Marquardt minimization, assuming Gaussian statistics (see \citealt{bekiaris2016}). We applied this fitting procedure to the observed cube, and determined the confidence intervals as in \rsec{SEC:flux_extract}: we created a set of simulated cubes by perturbing the best-fitting model with Gaussian noise, convolved them with the dirty beam, and applied the same fitting procedure to all the simulated cubes to obtain the distribution of best-fit values. The formal best-fit and residuals are shown on \rfigs{FIG:cutout} and \ref{FIG:disk_model_maps}.

We found the systemic redshift of \cplus is $z=3.7087 \pm 0.0004$, while the Balmer lines of \jekyll are at $z=3.7174 \pm 0.0009$. The corresponding proper velocity difference is $549\pm60\,\kms$, and is highly significant. Indeed, the uncertainty on the wavelength calibration of MOSFIRE is only $0.1,\AA$ or $1.3\,\kms$ \citep{nanayakkara2016}, and the observed frequency of ALMA is known by construction. In addition, both spectra were converted to the solar-system barycenter reference frame, and we used vacuum rest-wavelengths for both the Balmer and \cplus lines. The dominant source of uncertainty on the velocity offset is thus the statistical uncertainty quoted above.

The total line flux is $\fcplus=1.85\pm0.22\,\Jykms$, which translates into a luminosity of $\lcplus=(8.4\pm1.0)\times10^{8}\,\lsun$. The inclination is relatively low, $i=19$ to $55\degr$, while the turnover radius is essentially unresolved, $r_t = 0^{+0.16}_{-0}\,\kpc$. The half-light radius of the \cplus emission is consistent with being the same as that of the dust continuum: $0.11\pm0.03\arcsec$ or $0.80\pm0.24\,\kpc$. The disk is rotating rapidly, with a period of only $t_{\rm rot} = 8.4^{+7.9}_{-2.8}\,\Myr$ and a high velocity of $v_{2.2}=781^{+218}_{-366}\,\kms$. Consequently the inferred dynamical mass is also high: $\mdyn = 1.3^{+1.2}_{-0.8}\times10^{11}\,\msun$.

The \cplus-to-FIR ratio of $\log_{10}(\lcplus/\lfir) = -2.91^{+0.19}_{-0.13}$ is a factor $3.6\pm1.3$ lower than the normal value in the local Universe \citep{malhotra1997}, which clearly places this galaxy in the ``\cplus deficit'' regime (see \rfig{FIG:cplusfir}). This is discussed further in \rsec{SEC:cplus_deficit}.

\section{Stellar emission \label{SEC:obs:uvnir}}

\begin{figure}
\begin{center}
\includegraphics[width=0.46\textwidth]{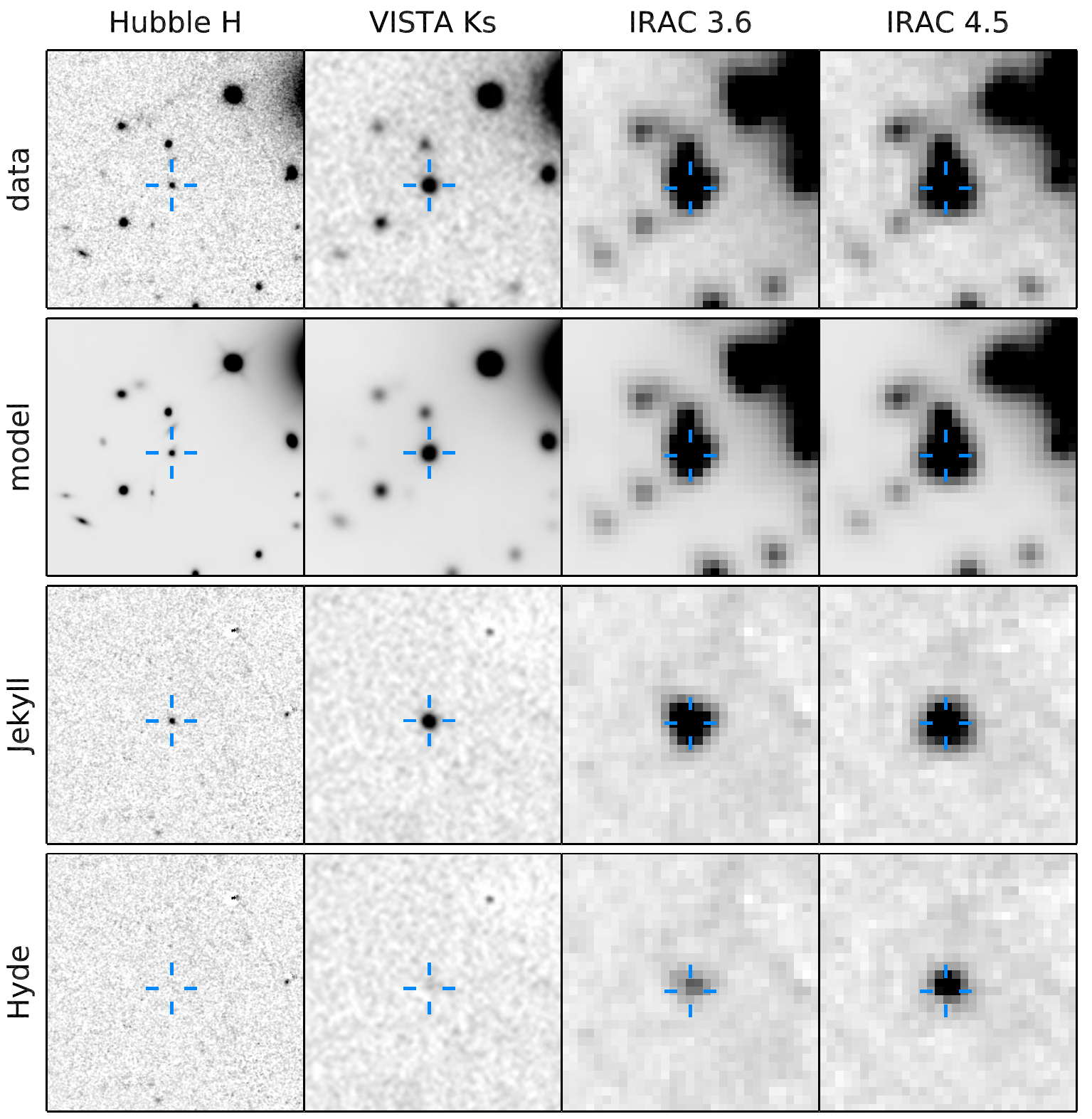}
\end{center}
\caption{Cutouts of the \hubble F160W, VISTA $K_{\rm s}$, IRAC $3.6$ and $4.5\,\um$ (from left to right). The first row shows the original images, the second row shows our best model, the third row shows all sources subtracted except \jekyll, while the fourth row shows the same thing for \hyde. Each image is $18\arcsec\times18\arcsec$, and the color table is the same for all images in a given column. The position of \jekyll is indicated with a blue cross. \label{FIG:stellar_cuts}}
\end{figure}

\subsection{Photometry \label{SEC:photometry}}

\begin{figure*}
\begin{center}
\includegraphics[width=\textwidth]{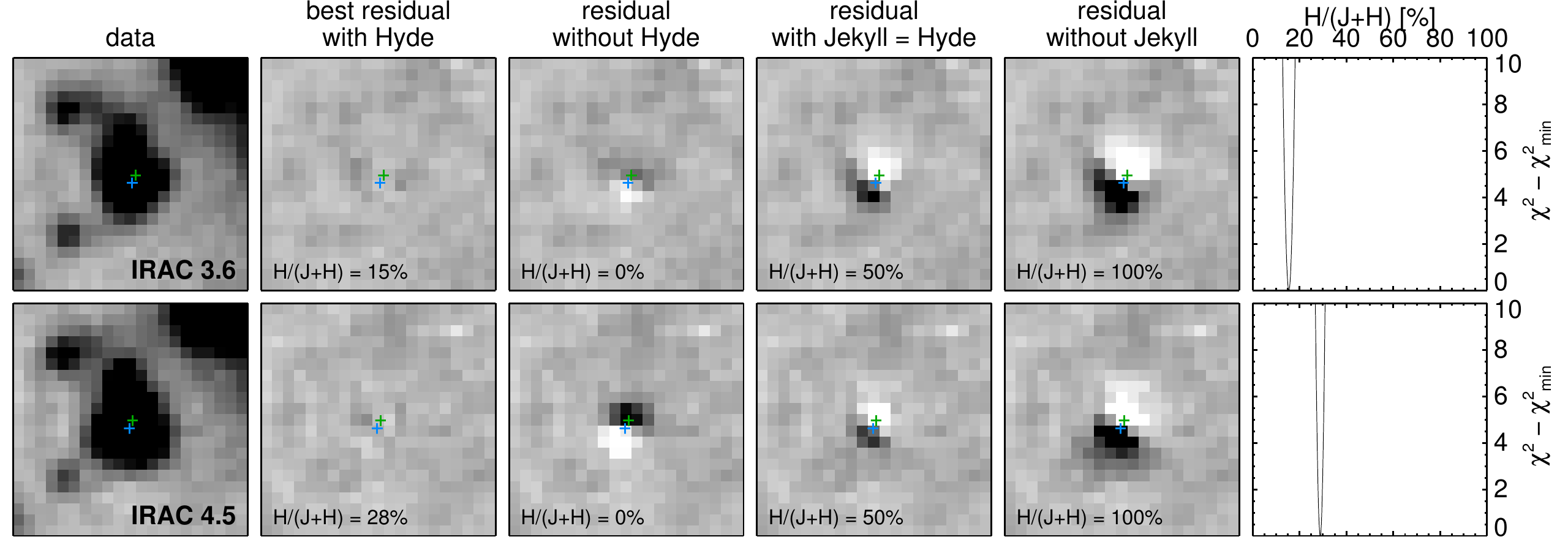}
\end{center}
\caption{Residuals of the \spitzer IRAC $3.6$ (top) and $4.5\,\um$ (bottom) images. From left to right: original image, best-fit residual, residual without \hyde, residual assuming the same flux for \jekyll and \hyde, residual without \jekyll, and $\chi^2$ of the fit as a function of the flux ratio $S_{\hyde}/(S_\jekyll + S_\hyde)$. Each cutout is $12\arcsec\times12\arcsec$, and the centroids of \jekyll and \hyde are shown with blue and green crosses, respectively. \label{FIG:irac_resid}}
\end{figure*}

Since \jekyll and \hyde are extremely close, we performed a careful deblending to see if we could detect the stellar emission of \hyde. We did this by modeling the profile of all galaxies within a radius of $15\arcsec$ with \galfit \citep{peng2002} on the \hubble F160W image, using single S\'ersic profiles of varying position, total flux, half-light radius, position angle and S\'ersic index. Since \hyde is not detected on the \hubble images, we assumed instead the disk profile obtained by modeling its dust emission (see \rsec{SEC:flux_extract}). We then used these profiles to build the models of all galaxies on the other bands using the appropriate point spread function (PSF), and fit the images as a linear combination of all these models plus a constant background (fluxes were allowed to be negative). Prior to the fit, the neighboring bright elliptical was modeled with four S\'ersic profiles, adjusted with all other sources masked (including a lensed galaxy close to the core of the elliptical), and was subtracted from each image. A star spike was also removed from the \hubble images. Using this method, we extracted fluxes on all the Subaru, \hubble, ZFOURGE, VISTA and \spitzer IRAC broad-band images, covering $\lambda=0.45$ to $8\,\um$. The result of this deblending depends on the assumption that the shape of all galaxies (including \jekyll) does not vary strongly between the \hst $H$ band and the other bands, in particular \spitzer IRAC. The clean residuals (see below and \rfig{FIG:stellar_cuts}) suggest that this is not a major issue.

To estimate uncertainties, we performed a Monte Carlo simulation where we varied the noise in each image by extracting a random portion of empty sky from the residual image, and co-adding it on top of \jekyll \& \hyde. This naturally accounts for correlated noise and large-scale background fluctuations. The PSFs were obtained by stacking stars in the neighborhood of our two galaxies, performing sub-pixel alignment using bicubic interpolation, except for \spitzer IRAC where we built a custom PSF by co-adding rotated version of the in-flight PSF matching the orientation of the telescope through the various AORs, weighted by their respective exposure time. \cite{labbe2015} showed that the IRAC PSF is very stable in time, such that this procedure produces very accurate PSFs that can be used to go deeper than the image's confusion limit. Photometric zero points were matched to that of ZFOURGE \citep{straatman2016}.

Obtaining an accurate de-blending of the \jekyll \& \hyde pair required not only an excellent knowledge of the PSF, but also of the astrometry. To ensure our astrometry was well matched, we slightly shifted the WCS coordinate system of all the images until no residual remained for all the bright sources surrounding our two galaxies (to avoid biasing our results, the residuals at the location of \jekyll \& \hyde were ignored in this process). These shifts were no larger than $0.05\,\arcsec$ for all bands but \spitzer IRAC, where they reached up to $0.1\,\arcsec$. Most importantly, we also randomly shifted the position of \hyde's model in the Monte Carlo simulations used to estimate flux uncertainties, using a Gaussian distribution and the relative astrometry accuracy between ALMA and \hubble quantified in \rsec{SEC:astrometry} ($\sim0.08\arcsec$). This step significantly increased the uncertainties in the \spitzer bands.

We could not validate the astrometry of the \spitzer IRAC $5.8$ and $8\,\um$ images, since the $S/N$ there is low and not enough sources are detected in the immediate neighborhood. For these bands we therefore only measured the total photometry of the \jekyll \& \hyde system. The flux of \jekyll was then subtracted from these values, by extrapolation of the best-fitting stellar template (see next section). The remaining flux was attributed entirely to \hyde.

The resulting residual images are displayed in \rfig{FIG:stellar_cuts}, and the fluxes are displayed in \rfig{FIG:seds}. We found that \hyde is clearly detected in the first two \spitzer IRAC channels ($[3.6]=23.7$ and $[4.5]=22.7$), barely detected in the \Ks band ($K_{\rm s} = 25.2$), and undetected in all the bluer bands, including those from \hubble ($3\sigma$ upper limit of $H>26.3$). This implies very red colors, $H-[4.5] = 5.1\pm0.8$, similar to that of ``$H$-dropout'' galaxies \citep{wang2016}, and strong attenuation by dust. We describe how we modeled this photometry in the next section and discuss the results of the modeling in \rsec{SEC:stellar_pop}.

Even accounting for the uncertainty in the relative astrometry between ALMA and \hst, the flux ratios between \jekyll and \hyde is well constrained. In the Monte Carlo simulations, the ratio $S_{\hyde}/(S_{\jekyll}+S_{\hyde})$ was $15^{+3}_{-2}\%$ and $28^{+6}_{-4}\%$ in the \spitzer $3.6$ and $4.5\,\um$ bands, respectively. Using a simpler $\chi^2$ approach (i.e., ignoring the uncertainty on the relative astrometry), we obtained instead $15.4\pm0.8\%$ and $28.8\pm0.6\%$ (see \rfig{FIG:irac_resid}, rightmost panel). The residuals obtained by fixing the flux ratio of \jekyll and \hyde to $0$, $50$ and $100\%$ are shown in \rfig{FIG:irac_resid}, and clearly show that either of these assumptions provides a poor fit compared to our best values of $15$ and $28\%$. This demonstrates that \hyde is required to fit the IRAC emission, and that it cannot be brighter than \jekyll.

Lastly, we have also tried to fit the \hst and \spitzer IRAC $4.5\,\um$ images by freely varying the centroids (and for \hst only, the profile shapes) of both \jekyll and \hyde. These fits therefore did not make use of \hyde's centroid as observed in the ALMA image. In the \hst image, we found that \hyde is offset from \jekyll by $\Delta\alpha=-0.11\pm0.05\arcsec$ and $\Delta\delta=+0.40\pm0.04\arcsec$, while in IRAC we found $\Delta\alpha=-0.047\pm0.02\arcsec$ and $\Delta\delta=+0.40\pm0.03\arcsec$. Both values are consistent with the ALMA position (offset of $0.02\pm0.08\arcsec$ and $0.09\pm0.06\arcsec$, respectively), which provides an independent evidence of \hyde's existence as a separate source.

\subsection{Modeling \label{SEC:stellar_fit}}

\subsubsection{Description of the code and key assumptions}

The photometry of both objects was modeled using FAST++\footnote{\url{https://github.com/cschreib/fastpp}}, a full rewrite of FAST \citep{kriek2009} that can handle much larger parameter grids and offers additional features. Among these new features is the ability to generate composite templates with any star-formation history (SFH) using a combination of \cite{bruzual2003} single stellar populations. A second important feature is the possibility to constrain the fit using a Gaussian prior on the infrared luminosity $\lir$, which can help pin down the correct amount of dust attenuation and improve the constraints on the other fit parameters. This code will be described in more detail in a separate paper (Schreiber et al.~in prep.), and we provide a brief summary here for completeness.

The $\lir$ predicted by a given model on the grid is computed as the bolometric luminosity absorbed by dust, i.e., the difference in luminosity before and after applying dust attenuation to the template spectrum, assuming the galaxy's flux is isotropic (see \citealt{charlot2000,dacunha2008,noll2009}). We thus used the values of $\lir$ determined in \rsec{SEC:dustprops} to further constrain the fit. Our adopted dust model is the same as that of FAST, and it assumes a uniform attenuation ($A_{\rm V}$) for the whole galaxy. This implies that dust is screening all stars uniformly, regardless of their age, which is usually a crude assumption. Here we argue that there is little room for differential attenuation, given the small sizes of \jekyll and \hyde (see \rsec{SEC:flux_extract}) and the necessarily short timescales involved in their formation. A uniform screen model is therefore a reasonable choice. Compared to models which assume lower attenuation for older stars, the $\lir$ predicted by our model will tend to include a larger proportion of energy from old-to-intermediate age stars, and consequently, at fixed $\lir$ our model will allow lower levels of on-going star-formation (see also \citealt{sklias2017}). This ``energy balance'' assumption has been shown to fail in strong starburst galaxies, possibly because of optically thick emission; these cases can be easily spotted as the model then provides a poor fit to the data \citep{sklias2017}. This did not happen here.

As in FAST, a ``template error function'' is added quadratically to the flux uncertainties, taking into account the uncertainty in the stellar population model (in practice, this prevents the $S/N$ in any single band from reaching values larger than $20$, see also \citealt{brammer2008}). This error function is not applied to the MOSFIRE spectrum of \jekyll. Instead, to reflect the fact that the relative flux between two spectral elements is more accurately known than their absolute flux, the code introduces an additional free normalization factor when fitting the spectrum. As a consequence, only the features of the spectrum contribute to the $\chi^2$ (i.e., the strength of the absorption lines), and not its integrated flux.

Finally, we did not include emission lines in the fit. While $z=3.7$ is the redshift where H$\alpha$ enters the IRAC $3.6\,\um$ band, possibly contributing significantly to the broad-band flux \citep[e.g.,][]{stark2013}, this is not a problem for our galaxies. Indeed, for \jekyll a contribution of more than $5\%$ of the IRAC flux would require $\sfr > 35\,\msun/\yr$, which is ruled out by the dust continuum and the absence of emission line in the \Ks-band spectrum \citep{glazebrook2017}. For \hyde, the modeling without emission line suggests $A_{\rm V} = 3.5\,{\rm mag}$ (see \rsec{SEC:stellar_pop}), therefore a $>5\%$ contribution of the $3.6\,\um$ flux would require $\sfr > 170\,\msun/\yr$, which is higher than that inferred from the infrared luminosity. The possibility of substantial contamination of the $3.6\,\um$ band can thus be safely ignored here. The remaining potential contaminant is [\ion{O}{iii}], which could contribute to the \Ks band flux. Because of the mask design, the MOSFIRE spectrum of \jekyll used by Glazebrook et al.~did not cover this line. However, this system was later re-observed as filler in the MOSEL program (Tran et al.~in prep.), with a 1.6 hours exposure in K and a different wavelength coverage including [\ion{O}{iii}]. No line was found in this new spectrum, and since the $0.7\arcsec$ slit is wide enough to include potential emission lines from \hyde as well, we confidently ignored strong emission lines in this analysis.

\subsubsection{The grid}

\begin{figure}
\begin{center}
\includegraphics[width=0.5\textwidth]{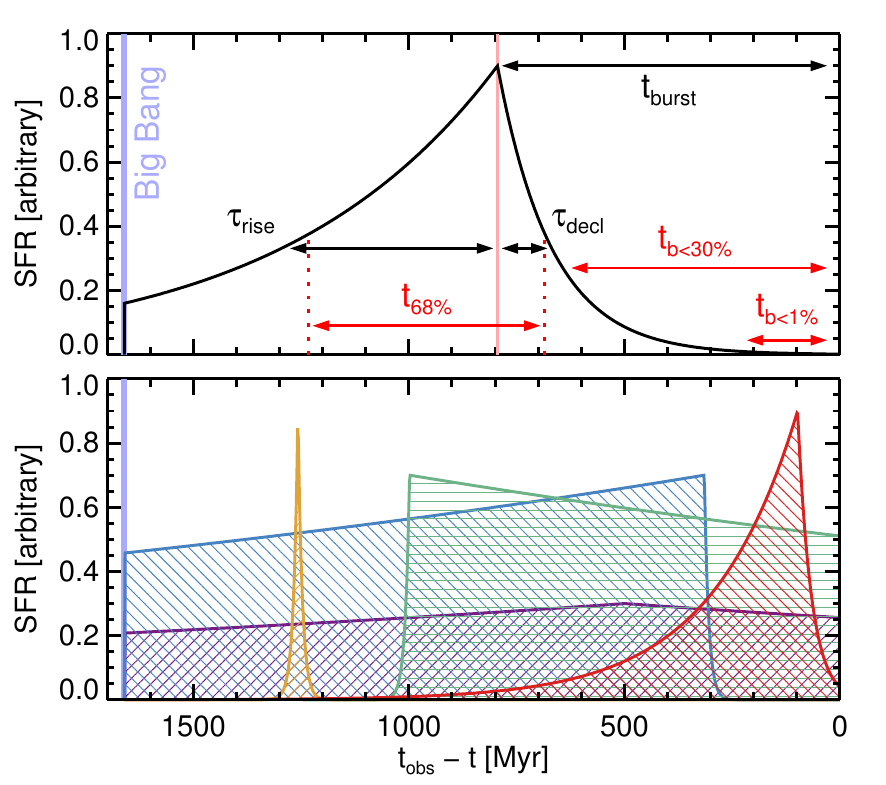}
\end{center}
\caption{Cartoon picture (top) and examples (bottom) of our model star-formation history, i.e., the $\sfr$ as a function of $t_{\rm obs}-t$, where $t_{\rm obs}$ is the time of observation ($z\sim3.7$). In the top row, we show the parameters of the model SFH in black, and the post-processed quantities in red. The examples shown in the bottom row are a roughly constant SFH since the Big Bang (purple), a roughly constant SFH starting $1\,\Gyr$ ago (green), a roughly constant SFH with an abrupt quenching $300\,\Myr$ ago (blue), a brief and old burst (yellow), and a slowly rising SFH with a recent decline (red). Many more combinations are possible but not shown for clarity. \label{FIG:sfh}}
\end{figure}

Fixing the redshifts to their spectroscopic values, we modeled the two galaxies using a ``double-$\tau$'' SFH, i.e., an exponential rise followed by an exponential decline (see \rfig{FIG:sfh}). Compared to the top-hat SFH used in \cite{glazebrook2017}, this parametrization allows additional scenarios where the $\sfr$ is reduced gradually over time, rather than being abruptly truncated. The two phases can have different $e$-folding times, $\tau_{\rm rise}$ and $\tau_{\rm decl}$, respectively. The corresponding analytic expression is
\begin{align}
\sfr(t) = C\times\left\{\begin{array}{ll}
e^{(t_{\rm burst} - t)/\tau_{\rm rise}} & \text{for $t > t_{\rm burst}$,} \\
e^{(t - t_{\rm burst})/\tau_{\rm decl}} & \text{for $t \le t_{\rm burst}$,} \\
\end{array}\right.\label{EQ:sfh}
\end{align}
where $t$ is the lookback time. The time of peak $\sfr$, $t_{\rm burst}$, was varied from $10\,\Myr$ to the age of the Universe ($1.65\,\Gyr$) in logarithmic steps of $0.05\,\dex$. The two $e$-folding times, $\tau_{\rm rise}$ and $\tau_{\rm decl}$, were varied from $10\,\Myr$ to $3\,\Gyr$ in steps of $0.1\,\dex$. The constant $C$, which can be identified as the peak $\sfr$, was finally used to adjust the normalization of the SFH for each combination of the above parameters, and eventually determined other derived properties such as the stellar mass. For each SFH, we computed the average $\sfr$ over the last $10$ and $100\,\Myr$ ($\sfr_{\rm 10\,\Myr}$ and $\sfr_{\rm 100\,\Myr}$, respectively). In the following we refer to the ``current'' $\sfr$ as the average of the last $10\,\Myr$, since variations of the $\sfr$ on shorter timescales are not constrained by the photometry; this average is thus better measured than the instantaneous $\sfr$ one would derive from \req{EQ:sfh}.

The parameters $t_{\rm burst}$, $\tau_{\rm rise}$ and $\tau_{\rm decl}$ were chosen to span a wide range of SFHs (as demonstrated in \rfig{FIG:sfh}). However, their physical interpretation is not immediate, and the resulting parameter space contains some degeneracies. For example, the value of $\tau_{\rm decl}$ is mostly irrelevant when $t_{\rm burst}$ is very small, and conversely the value of $t_{\rm burst}$ is also irrelevant when both $e$-folding times are large. We therefore post-processed the resulting SFHs to define a handful of well-behaved quantities. First, defining $b = \sfr(t)/\sfr_{\rm max}$ as the ratio between the instantaneous and maximum $\sfr$, we computed the time spent with $b < 1\%$ and $b < 30\%$ starting from the epoch of observation and running backwards in time ($t_{b<1\%}$ and $t_{b<30\%}$, respectively). This can be identified as the duration of quiescence ($t_{\rm qu}$), and will be equal to zero by definition if the galaxy is not quiescent at the time of observation. Second, we computed the shortest time interval over which $68$ and $95\%$ of the star-formation happened ($t_{68\%}$ and $t_{95\%}$, respectively), which can be identified as the formation timescale ($t_{\rm sf}$). These quantities are illustrated in \rfig{FIG:sfh}. Finally, to locate the main star-forming epoch, we computed the $\sfr$-weighted lookback time $t_{\rm form} = \int\!t\,\sfr(t)\,\dd t / \int\!\sfr(t)\,\dd t$ and the associated redshift $z_{\rm form}$.

We then varied the attenuation from $A_{\rm V} = 0$ to $6$ magnitudes (assuming the \citealt{calzetti2000} absorption curve), and fixed the metallicity to the solar value (leaving it free had a negligible impact on the best fit values). A total of about $2$ million models were generated and compared to the photometry of both galaxies. For \jekyll we also included the MOSFIRE spectrum, coarsely binned to avoid having to accurately reproduce the velocity dispersion of the absorption lines; in practice this amounts to introducing a prior on the Balmer equivalent widths. This resulted in $25$ and $20$ degrees of freedom for \jekyll and \hyde, respectively. Finally, confidence intervals were derived from the minimum and maximum values allowed in the volume of the grid with $\chi^2 - \chi^2_{\rm min} < 2.71$ (i.e., these are $90\%$ confidence intervals; \citealt{avni1976}). As a cross check, we also performed 1000 Monte Carlo simulations where the photometry of each galaxy was perturbed within the estimated uncertainties and fit as the observed photometry, and we then computed the 5th and 95th percentiles of the parameter distributions. The resulting constraints on the fit parameters were similar but slightly tighter than those obtained using the $\chi^2$ criterion above; in order to be most conservative we used $\chi^2$-based confidence intervals throughout.

Using simulated bursty SFHs, we show in \rapp{APP:sfh} that the resulting constraints on the quenching and formation timescales are accurate even if the true SFH deviates from the ideal model of \req{EQ:sfh}. The only exception is when a second burst happened in the very early history of the galaxy. In these cases, two outcomes are possible: either the older burst is outshined by the latest burst and is thus mostly ignored (see also \citealt{papovich2001}), leading to underestimated stellar masses and formation timescales, or the fit to the photometry is visibly poor, with discrepancies of more than $2\sigma$ in the NIR bands. On no occasion was a star-forming SFH misclassified as quiescent, instead small residual $\sfr$s were found to potentially bias the quenching times to lower values.

Finally, we have tried to fit a more complex model than \req{EQ:sfh} to our galaxies by including a late exponentially rising burst active at the moment of observation, of variable intensity and $e$-folding time. The constraints for \jekyll were unchanged, and the only difference for \hyde was that additional solutions were allowed where the bulk of the galaxy formed very early ($z>5$) in a short burst, while the observed FIR emission was produced by a more recent burst of lower SFR $\sim 80\,\msun/\yr$. These solutions appear unrealistic: the main burst of star-formation would have happened earlier than in \jekyll and yet the galaxy would still contain more dust than \jekyll. Given that this additional complexity did not provide further useful information but introduced unrealistic scenarios, we decided to keep the simpler SFH of \req{EQ:sfh}.

\subsection{Results \label{SEC:stellar_pop}}

\begin{figure*}
\begin{center}
\includegraphics[width=0.49\textwidth]{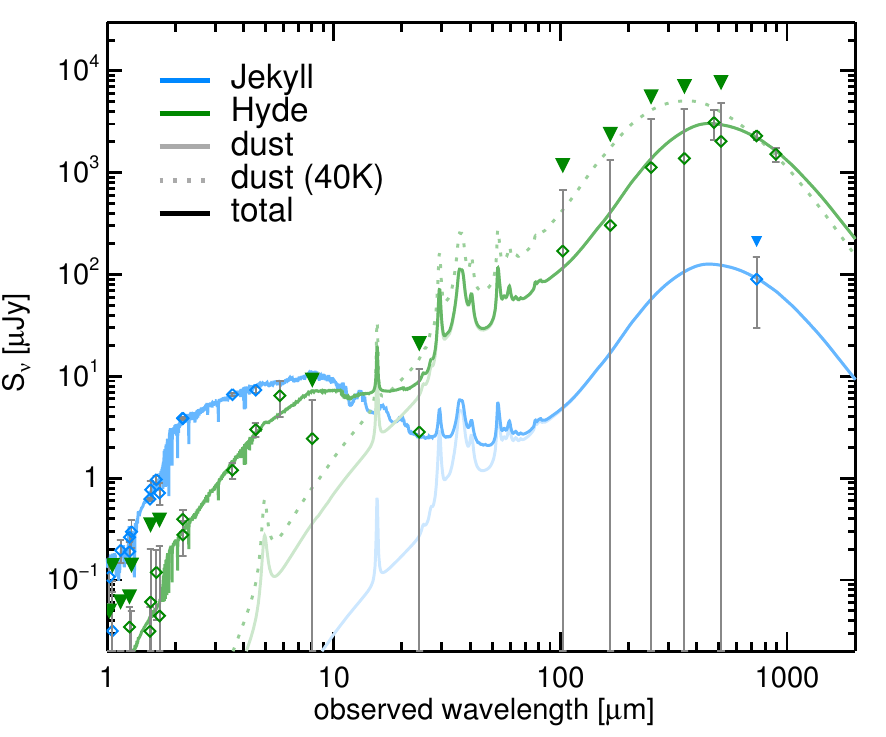}
\includegraphics[width=0.49\textwidth]{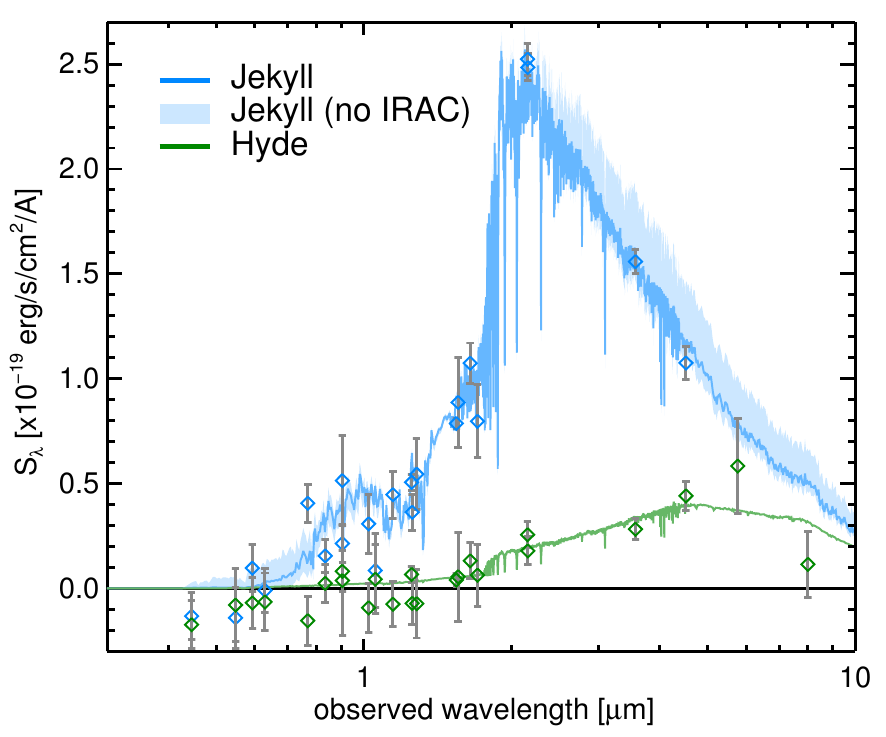}
\end{center}
\caption{{\bf Left:} Photometry of \jekyll (blue) and \hyde (green) from the UV to the sub-milimeter. The observed photometry is shown with diamonds (downward pointing triangles indicate $2\sigma$ upper limits for measurements of significance less than $2\sigma$). The best fitting dust model is shown with an pale line, and the total model (dust and stars) is shown with a darker line. For illustration, for \hyde we also show a dust model assuming $\tdust=40\,\kelvin$, which overpredicts the MIPS, Herschel and SCUBA fluxes. The dust model for \jekyll is only illustrative, and was simply normalized to match the constraint from the $744\,\um$ flux. {\bf Right:} Zoom-in on the stellar emission, shown in $S_\lambda$ instead of $S_\nu$. As described in the text, the $5.8$ and $8\,\um$ photometry are shown here only for \hyde; the fluxes in these bands were obtained from aperture photometry of the whole system, with the predicted contribution of \jekyll subtracted. Here we also show in light blue the range of possible SEDs for \jekyll when all the IRAC photometry is ignored in the fit. \label{FIG:seds}}
\end{figure*}

\begin{table}
\begin{center}
\caption{Properties of \jekyll \& \hyde. \label{TAB:props}}
\begin{tabular}{lcc}
\hline \\[-0.25cm]
                                  & \jekyll & \hyde \\ \hline\hline \\[-0.25cm]
$z_{\rm spec}$                    & $3.7174 \pm 0.0009$ & $3.7087 \pm 0.0004$ \\[0.15cm]
\multicolumn{3}{l}{Dust properties} \\\hline\\[-0.25cm]
$S_{744\um}$ ($\mJy$)             & $0.09\pm0.06$ & $2.31\pm0.14$ \\
$r_{\rm dust}$ ($\kpc$)           & -- & $0.67\pm0.14$ \\
$\tdust$ ($\kelvin$)              & $20$ -- $35$ $^a$ & $31^{+3}_{-4}$ \\[+0.05cm]
$\lir$ ($10^{12}\,\lsun$) $^b$    & $0.036^{+0.031}_{-0.024}$ & $1.1^{+0.4}_{-0.3}$ \\[+0.05cm]
$\lfir$ ($10^{12}\,\lsun$) $^c$   & $0.020^{+0.020}_{-0.014}$ & $0.67^{+0.25}_{-0.22}$ \\[+0.05cm]
$\sfr_{\rm IR}$ ($\msun/\yr$)     & $3.6^{+3.1}_{-2.4}$ & $110^{+43}_{-33}$ \\[+0.05cm]
$\mdust$ ($10^{8}\,\msun$) $^d$   & $0.19^{+0.26}_{-0.13}$ & $3.2^{+2.2}_{-1.0}$ \\[0.15cm]
$\Sigma_{\rm FIR}$ ($10^{11}\,\lsun/\kpc^2$) & -- & $2.3^{+1.9}_{-1.0}$ \\[0.15cm]
\multicolumn{3}{l}{\cplus properties} \\\hline\\[-0.25cm]
$S_{\cplus}$ ($\Jykms$)           & -- & $1.85\pm0.22$ \\
$r_{\rm \cplus}$ ($\kpc$)         & -- & $0.80\pm0.24$ \\
$\lcplus$ ($10^{8}\,\lsun$)       & -- & $8.4\pm1.0$ \\
$\log_{10}(\lcplus/\lfir)$        & -- & $-2.91^{+0.19}_{-0.13}$ \\[+0.05cm]
$v_{2.2}$ ($\kms$)                & -- & $781^{+218}_{-366}$ \\[+0.05cm]
$\sigma_v$ ($\kms$)               & -- & $79^{+30}_{-37}$ \\[+0.05cm]T
$v_{2.2}/\sigma_v$                & -- & $9.8^{+6.3}_{-4.6}$ \\[+0.05cm]
$t_{\rm rot}$ ($\Myr$)            & -- & $8.4^{+7.9}_{-2.8}$ \\[+0.05cm]
$\mdyn$ ($10^{11}\,\msun$)        & -- & $1.3^{+1.2}_{-0.8}$ \\[0.15cm]
\multicolumn{3}{l}{Inferred gas properties} \\\hline\\[-0.25cm]
$\mgas$ ($10^{10}\,\msun$)        & $<3.5$ & $3.6^{+4.3}_{-1.9}$ \\[0.15cm]
$\fgas$                           & $<25\%$ & $12$ -- $70\%$ \\[0.15cm]
$\Sigma_{\rm gas}$ ($10^{4}\,\msun/\pc^2$) & -- & $1.0^{+1.8}_{-0.6}$ \\[0.15cm]
$t_{\rm ff}$ ($\Myr$)             & -- & $1.2^{+0.9}_{-0.6}$ \\[+0.05cm]
\multicolumn{3}{l}{Stellar properties ($90\%$ confidence intervals)} \\\hline\\[-0.25cm]
$\mstar$ ($10^{11}\,\msun$)       & $1.03$ -- $1.35$ & $0.34$ -- $1.28$ \\
$r_\ast$ ($\kpc$)                 & $0.49\pm0.12$ $^e$ & -- \\
$\Sigma_\ast$ ($10^{4}\,\msun/\pc^2$) & $2.2$ -- $9.7$ & $0.9$ -- $6.6$ $^f$ \\
$A_{\rm V}$ (mag)                 & $0.19$ -- $0.48$ & $2.68$ -- $3.81$ \\
$\sfr_{10\,\Myr}$ ($\msun/\yr$)   & $0$ -- $0.48$ & $0$ -- $119$ \\
$\sfr_{10\,\Myr}/\sfrms$          & $0$ -- $10^{-3}$ & $0$ -- $0.72$ \\
$\sfr_{100\,\Myr}$ ($\msun/\yr$)  & $0$ -- $0.65$ & $0$ -- $828$ \\
$\Sigma_\sfr$ ($\msun/\yr/\kpc^2$) & $0$ -- $0.18s$ & $0$ -- $61$ \\
$t_{b<1\%}$ ($\Myr$)              & $210$ -- $661$ & $0$ -- $204$ \\
$t_{b<30\%}$ ($\Myr$)             & $337$ -- $724$ & $0$ -- $570$ \\
$t_{68\%}$ ($\Myr$)               & $22$ $^g$ -- $839$ & $22$ $^g$ -- $1079$ \\
$t_{95\%}$ ($\Myr$)               & $58$ $^g$ -- $1246$ & $58$ $^g$ -- $1566$ \\[0.1cm]
\hline
\end{tabular}
\end{center}
{\footnotesize $^a$ This is the range of temperatures assumed to estimate $\lir$ and $\mdust$ for \jekyll only. It is not a measurement. $^b$ $8$ to $1000\,\um$. $^c$ $42.5$ to $122.5\,\um$. $^d$ These dust masses correspond to a model with amorphous carbon grains, which provides values a factor $2.6$ lower than the graphite-based models commonly used in the literature (e.g., \citealt{draine2007}). $^e$ From \cite{straatman2015}, $68\%$ confidence interval. $^f$ Assuming that stars follow the same profile as dust. $^g$ These values are limited by the minimum $e$-folding times allowed in the grid.}
\end{table}

The results of the UV-to-FIR SED modeling (\rsec{SEC:stellar_fit}) are listed in \rtab{TAB:props} and illustrated in \rfig{FIG:tqtsf}.

\subsubsection{\jekyll}

We recovered the result of \cite{glazebrook2017}, namely that \jekyll has quenched $>$$210\,\Myr$ before we observed it, at $z\sim5$, with a formation redshift between $z_{\rm form}=5.4$ and $7.6$. The sum of the quenched and star-forming epochs leads to a total age of $t_{b<30\%} + t_{68\%} = 610\,\Myr$ to $1.1\,\Gyr$, which is slightly older than found by Glazebrook et al. Since some of the flux is now attributed to \hyde, the stellar mass of \jekyll has decreased by 30\% (-0.14\,\dex) compared to its initial estimation. The constraint from the observed $\lir$ rules out solutions with $A_{\rm V} > 0.5$ for \jekyll and tends to push the formation timescale toward larger values, albeit still within the error bars quoted by Glazebrook et al. These changes are not sufficient to erase the tension with galaxy formation models, as the baryon conversion efficiency for a formation at $z=5$ is still high ($60\%$). Therefore the conclusions presented in Glazebrook et al.~still apply.

We note that we reached this result even when we excluded the IRAC photometry from the fit; ignoring the IRAC fluxes would allow a larger stellar mass of up to $1.7\times10^{11}\,\msun$, but it would not impact the minimum mass. The rest of the data (i.e., mostly the \Ks-band flux, $H-K_{\rm s}$ color, $\lir$ limit, and MOSFIRE spectrum) indeed independently constrain the mass and SFH, and are sufficient to predict the IRAC $3.6$ and $4.5\,\um$ fluxes of \jekyll with an accuracy of $24$ and $28\%$, respectively (see \rfig{FIG:seds}, right). These results are thus insensitive to systematics in the IRAC de-blending. In addition, fitting only the $U$-to-\Ks broadband photometry also leads to a lower limit on the mass of $1.0\times10^{11}\,\msun$; then, the SFH becomes poorly constrained and the photometry allows dusty star-forming solutions with very extreme $M/L$, such that the maximum allowed mass increases to $2.9\times10^{12}\,\msun$. This shows that the red $H-K_{\rm s}$ color alone places a stringent and secure lower limit on the $M/L$ and the mass, since neither the $H$ nor the \Ks bands are significantly contaminated by \hyde.

A similar analysis of this galaxy pair was done in \cite{simpson2017}; they found a substantially lower mass of $0.8\times10^{11}\,\msun$ for \jekyll, which is below our minimum allowed mass. We attribute the source of this difference to the different UV-IR SED used for \hyde: using an average SMG SED and rescaling it to the observed ALMA flux for \hyde, Simpson et al.~estimated a contamination of $30\%$ to the \Ks band (they predicted a flux of $\sim1\,\uJy$). Instead, our explicit de-blending of the images showed that this value is only $6\%$ ($0.33\pm0.08\,\uJy$); the ZFOURGE \Ks-band has excellent spatial resolution ($0.47\arcsec$ FWHM), such that a $30\%$ contribution to the flux would be readily apparent (e.g., \rfig{FIG:stellar_cuts}). Their adopted SED also produces a higher contribution to the flux in the IRAC bands, albeit to a lesser extent. \Citet{dacunha2015} showed that the rest-optical fluxes of SMGs spans two orders of magnitude at fixed sub-mm flux (see their Figure 13), which implies that a simple rescaling of the average SMG SED cannot predict accurate optical fluxes; an explicit de-blending and SED fit, as used here, is needed for accurate stellar masses.

\begin{figure*}
\begin{center}
\includegraphics[width=0.85\textwidth]{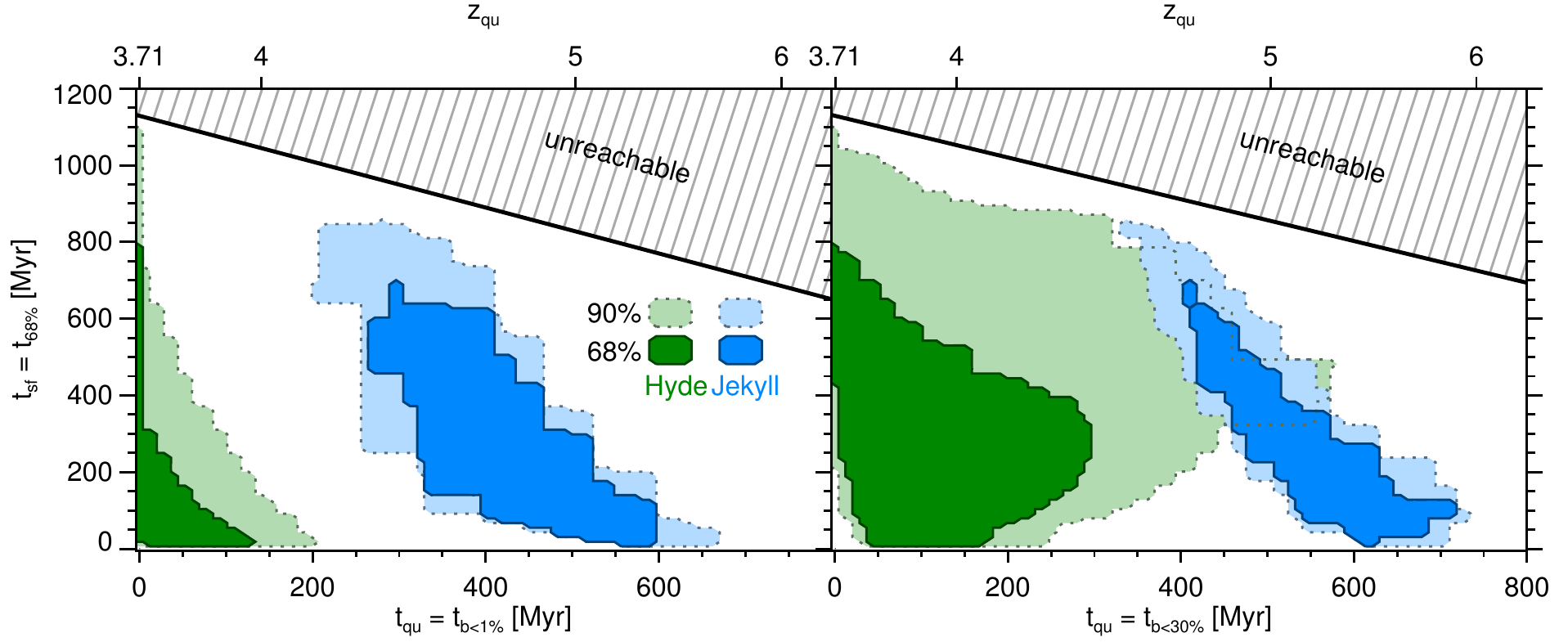}
\end{center}
\caption{Range of allowed values for the star-formation timescale ($t_{\rm sf}$), defined as the time over which $68\%$ of the star-formation happened, and the quiescence time ($t_{\rm qu}$), defined either as the time spent with less than $1\%$ (left) or $30\%$ (right) of the peak $\sfr$. The redshift at which the galaxy ``quenched'' is given on the top axis. The parameter spaces allowed for \jekyll and \hyde are shown in blue and green, respectively. The dark and light colored regions show the $68\%$ and $95\%$ confidence regions, respectively. The hashed region at the top indicates the part of the parameter space that would imply a formation before the Big Bang; such solutions were not explored. \label{FIG:tqtsf}}
\end{figure*}

\subsubsection{\hyde}

For \hyde, we found a large stellar mass either comparable to that of \jekyll or up to a factor three smaller, and a strong attenuation ($A_{\rm V} \sim 3.5\,{\rm mag}$) which is substantially redder than the average SMG ($A_{\rm V} \sim 2$; \citealt{dacunha2015}). The constraints on the star-formation history are looser than for \jekyll, however they are far from devoid of information. In particular, the photometry allows scenarios where star-formation was quenched ($b<1\%$) up to $200\,\Myr$ prior to observation, and rules out current $\sfr$ higher than $120\,\msun/\yr$. In all the models allowed by the fit, the galaxy is located below the $z=4$ main sequence by at least a factor $1.4$ \citep{schreiber2017-b}. This includes scenarios where the galaxy is simply on the lower end of the main sequence (with a main sequence dispersion of $0.3\,\dex$, there is a $30\%$ chance of being located a factor $1.4$ below the fiducial main sequence locus) as well as scenarios where the galaxy has recently stopped forming stars. Indeed, the $\sfr$ averaged over the last $10$ or $100\,\Myr$ could also be zero, in which case the FIR emission in the model comes from obscured non-OB stars \citep[e.g.,][]{bendo2012,bendo2015,eufrasio2017}.

Other parameters like the formation timescale cover a broad range when marginalizing over the allowed parameter space. However, the allowed values span different ranges depending on whether \hyde has quenched or not (see \rfig{FIG:tqtsf}). For quenched models with $t_{b<1\%} > 50\,\Myr$, $t_{68\%}$ can be at most $450\,\Myr$ (and less than $150\,\Myr$ at $68\%$ confidence), and the current $\sfr < 10\,\msun/\yr$. On the other hand, if the galaxy is still forming stars ($t_{b<30\%} = 0$) the formation timescale must be at least $250\,\Myr$ and the $\sfr$ averaged over the last $100\,\Myr$ must be less than $200\,\msun/\yr$. Therefore, either the galaxy has quenched after a brief but intense star-formation episode, or it has continuously formed stars at moderate rates over longer timescales. As we discuss in \rsec{SEC:discussion}, the compactness of the galaxy and the deficit of \cplus emission favor the former hypothesis.

Finally we note that the observed $\lir$ of \hyde provides crucial constraints on its modeled star-formation history. If $\lir$ had not been used in the fit, the whole parameter space would have been degenerate, and both the quiescence time and the formation timescale would be unconstrained.

\section{Discussion \label{SEC:discussion}}

Using the diverse data and modeling presented in the previous sections, we now proceed to discuss the implications for the two galaxies studied in this paper.

While we were analyzing the data, \cite{simpson2017} concurrently performed a similar analysis to that undertaken here, but using the shallower ALMA data in which the sub-millimeter emission was first detected, and without the information of the \cplus emission. Assuming the sub-millimeter emission originates from an obscured component within the same galaxy, they subtracted this obscured component from the total photometry using an average optical SED for SMGs and re-evaluated the stellar mass of the quiescent component. They concluded that the mass reported in \cite{glazebrook2017} had been overestimated by a factor two or more, and that after correction the tension with models (e.g., \citealt{wellons2015,dave2016}) was erased. They further argued that sub-millimeter emission is not an unusual feature in so-called post-starburst galaxies, and implied that the galaxy may not be as old as it was initially claimed.

Based on the new ALMA data and an explicit de-blending of the UV-near-IR imaging, our findings are not consistent with those of \cite{simpson2017}. We obtained definite proof that the sub-millimeter emission is in fact produced by a separate galaxy (see \rsec{SEC:hyde_galaxy}), which is extremely obscured. The colors of the dusty galaxy are redder than assumed by Simpson et al., resulting in a lower contamination of the photometry of the quiescent galaxy and a milder reduction of its stellar mass (see the discussion in \rsec{SEC:stellar_pop}). The quiescent galaxy, in turn, is not detected in our deep dust continuum map, imposing a stringent upper limit on its obscured $\sfr$. We discuss this further in the next section.

\subsection{No significant dust-obscured star-formation in \jekyll \label{SEC:dusty_psb}}

\begin{figure*}
\begin{center}
\includegraphics[width=0.9\textwidth]{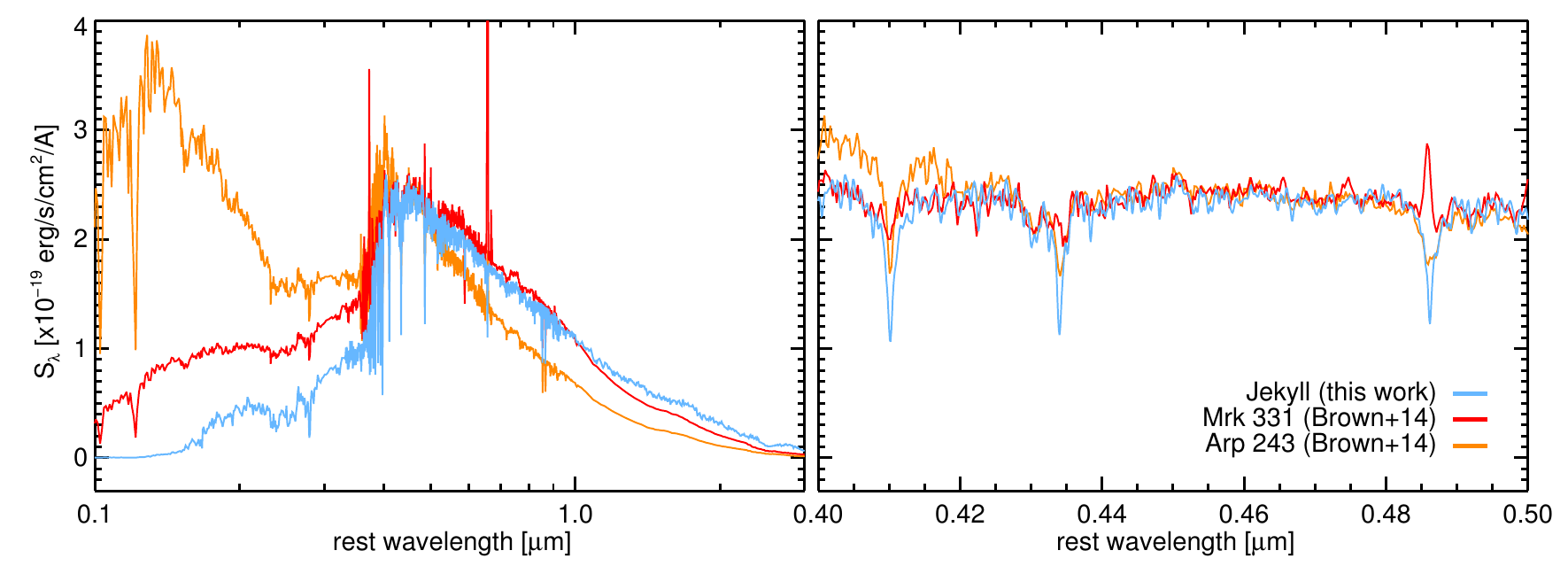}
\end{center}
\caption{Best stellar model of \jekyll, compared to the models of two local LIRGs with Balmer absorption lines, Mrk 331 and Arp 224, obtained by \cite{brown2014}. All these models use spectra to constrain the strength of the Balmer absorption lines (the spectra are not shown for clarity). the models of the two LIRGs were rescaled to match the continuum level of \jekyll at rest wavelengths between $0.45$ and $0.48\,\um$. \label{FIG:poggianti}}
\end{figure*}

\cite{simpson2017} argued that deep Balmer absorption lines, as observed in \jekyll, are not uniquely associated with truly post-starburst galaxies and can be observed in dusty starbursts as well. This can happen if the A stars, responsible for the Balmer absorption features, have escaped the dust clouds, where star-formation is still on-going and fully obscured. Such galaxies are labeled ``e(a)'' \citep{poggianti2000}. Simpson et al.~quoted Mrk 331 as an example.

We have shown in \rsec{SEC:flux_extract} that there is no detectable sub-millimeter emission at the position of \jekyll, therefore the amount of obscured star-formation in this galaxy must be particularly small ($\sfr_{\rm IR} < 13\,\msun/\yr$ at $3\sigma$, converting the limit on the observed $\lir$ to $\sfr$ directly, assuming no contribution of older stars to the dust heating). In the following, we nevertheless argue that \jekyll has very different spectral properties than those ``e(a)'' galaxies, and therefore the possibility of it belonging to this class of object could have been discarded from the start.

We display our best model for \jekyll and that of Mrk 331 as obtained by \cite{brown2014} in \rfig{FIG:poggianti}. It is immediately apparent that Mrk 331 has a weaker Balmer break, implying a younger stellar population. But more importantly it has an H$\delta$ equivalent width of only $4.1$\,\AA, a factor two lower than that observed in \jekyll, and H$\beta$ in emission rather than absorption (see \citealt{poggianti2000}). It is thus clear that Mrk 331 is not a good analog of \jekyll.

\cite{poggianti2000} analyzed the Balmer equivalent widths of a complete sample of luminous infrared galaxies ($\lir > 3\times10^{11}\,\lsun$) drawn from the IRAS $2\,{\rm Jy}$ catalog (see \citealt{wu1998-a}). This catalog covers $35\,000$ square degrees with redshifts up to $z\sim0.1$, which corresponds to a volume $300$ times larger than that covered by ZFOURGE at $3.4<z<4.2$. Of the 52 galaxies with spectral coverage for both H$\beta$ and H$\delta$ (60\% of their sample), none has ${\rm EW_{H\delta}} > 7\,\AA$ and ${\rm EW_{H\beta}} > 7\,\AA$, while \jekyll has ${\rm EW_{H\delta}} = 9.8\pm2.6\,\AA$ and ${\rm EW_{H\beta}} = 19.2\pm4.2\,\AA$ (NB: in their Table 1, Poggianti et al.~listed the equivalent widths of H$\delta$ with positive values for absorption, but they used the opposite convention for H$\beta$). The closest match is Arp 243 (IRAS 08354+2555), with ${\rm EW_{H\delta}} = 7.2\,\AA$ and ${\rm EW_{H\beta}} = 5.3\,\AA$, which we also show on \rfig{FIG:poggianti}. While the absorption lines are stronger than in Mrk 331, the Balmer break is also much weaker.

Despite the larger volume of the IRAS catalog, no galaxy from this sample matches simultaneously the strong Balmer break, H$\delta$, and H$\beta$ absorption observed in \jekyll. It is therefore clear that \jekyll has little in common with ``e(a)'' galaxies, and its non-detection on our deep dust continuum map confirms this conclusion.

\subsection{\hyde is a separate galaxy \label{SEC:hyde_galaxy}}

Given the close proximity of \jekyll and \hyde, it is legitimate to wonder if these are, indeed, two separate galaxies or two components of a single galaxy. This distinction goes beyond mere semantics: if these are two different galaxies, their formation history can be studied separately as their stellar, dust, and gas component have never mixed. They can be considered as two closed boxes with no exchange of matter. On the other hand, if these were two regions of a single galaxy, it would be possible for matter to migrate from one region to the other, and only the summed star-formation history of both components would be meaningful. One could imagine, for example, that the entire galaxy has been forming stars continuously, and that old stars have migrated out of the dusty star-forming region a few hundred million years prior to observation.

The answer to this question therefore determines whether or not we have found a truly quiescent galaxy at $z\sim4$. We stress however that there is one fact that holds regardless: the detection of the Balmer absorption lines in \jekyll imposes, without a doubt, that about $10^{11}\,\msun$ of stars were already formed at $z\sim5$. The implied past $\sfr$ and its consequence on galaxy evolution models (see \rsec{SEC:stellar_pop} and \citealt{glazebrook2017}) is not changed by this discussion.

Based on the data we present in this paper, a number of arguments can be put forward to show that indeed these are two separate galaxies. First, the large line-of-sight velocity difference of $\sim$$550\,\kms$ demonstrates their existence as two kinematically separate components, rather than an homogeneous mixture of stars and dust. Second, the projected distance between \jekyll and \hyde corresponds to five times their respective half-light radii, which rules out the interpretation of this system as a smooth galaxy with an attenuation gradient. Indeed, while \cite{chen2015} showed that physical offsets as large as $\Delta p = 3.3\,\kpc$ are common when comparing the ALMA and \hst emission of $z\sim2$ SMGs, if caused by an attenuation gradient the amplitude of such offsets must naturally scale with a galaxy's size. Chen et al.~found an average stellar half-light radius of $r_{1/2}\sim4\,\kpc$ for their SMGs, implying an average $\Delta p/r_{1/2} \sim 0.8$. For a galaxy as small as \jekyll, this corresponds to a potential offset of the order of $0.4\,\kpc$ only, or $0.06\arcsec$, which is much smaller than the observed $0.43\arcsec$.

Third, the fact that their stellar masses are comparable rules out the possibility of \jekyll being a satellite clump in the disk of \hyde. This hypothesis could be suggested by the fact that low-mass UV-bright clumps are often found in the outskirts of SMGs (e.g., \citealt{targett2013}). Yet, beside its small size \jekyll has little in common with these clumps (see \citealt{guo2012}): it is massive and old, and dominates the flux at all $\lambda \leq 4.5\,\um$. In addition, the projected velocity predicted by our disk modeling at the position of \jekyll is $+289^{+54}_{-72}\,\kms$, which is only half of the observed velocity offset of $+549\pm60\,\kms$. Therefore, \jekyll cannot be part of \hyde's disk.

The fourth and last evidence that these are two separate galaxies lies in the velocity structure of the \cplus line. Indeed, ``double horn'' velocity profiles like that shown on \rfig{FIG:cutout} can only be obtained with a flattened rotation curve, which implies that the \cplus emission is confined within its own dark matter halo (we tried building a model with a clump embedded in the halo of \jekyll, but this never produced such double-horn profile). Linear velocity gradients on scales larger than $0.6\,\kpc$ ($0.1\arcsec$) are ruled out by our disk modeling. A similar velocity profile could also be produced by two dispersion-dominated components of equal mass but different systemic velocity, e.g., an on-going merger of two dusty galaxies, but there is no evidence that the dust continuum emission has two spatial components. Ruling out this possibility would require a spectrum with a higher $S/N$ than we have here. Finally, the interpretation of the \cplus emission generated by outflowing material from \jekyll is ruled out by the detection of dust and stellar continuum spatially-coincident with the line emission.

Given this suite of evidence, the hypothesis of this \hubble and ALMA emission coming from a single galaxy appears unlikely. We thus conclude that \jekyll and \hyde are indeed two separate galaxies, and therefore that \jekyll is a galaxy in which star-formation has uniformly stopped sometime around $z\sim5$.

Incidentally, the spectroscopic redshift of \hyde constitutes one of the few robust redshift measurement of an ``$H$-dropout'' galaxy (see also \citealt{daddi2009} for GN-10 at $z=4.04$ and \cite{walter2012} for HDF-850.1 at $z=5.18$). The $H$-dropout population was first identified in the \spitzer IRAC images as sources having no counter-part in the deep $H$ or $K$-band images, implying high redshifts, large stellar masses and extreme obscuration (e.g., $H - [4.5] > 2.5$, \citealt{huang2011,caputi2015,wang2016}). This obscuration makes it impossible to determine redshifts using nebular lines, and their SEDs are also lacking identifiable features such as the Lyman or Balmer break so their photometric redshifts are poorly constrained. \cite{wang2016} showed that, if all at $z\sim4$, these galaxies could contribute significantly to the mass function and cosmic $\sfr$ density, but were previously missing from most high redshift census. The confirmation of \hyde at $z\sim4$ supports this result and highlights the importance of better understanding this population.

\subsection{Compactness as a tracer of quenching \label{SEC:size}}

\begin{figure}
\begin{center}
\includegraphics[width=0.45\textwidth]{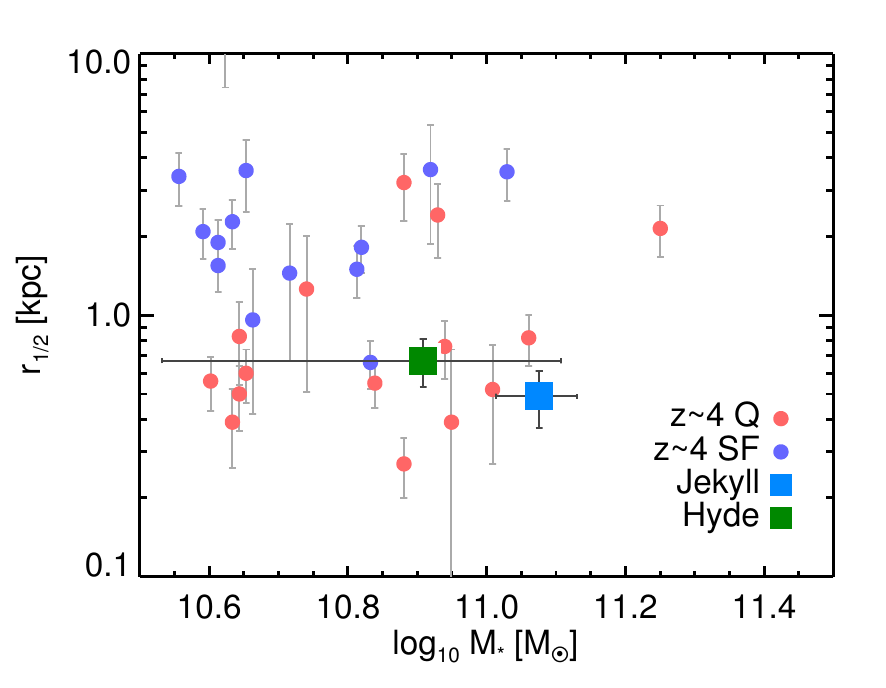}
\end{center}
\caption{Relation between the half-light radius ($r_{1/2}$) and the stellar mass for ZFOURGE galaxies with \hst coverage (80\% of the whole sample). \jekyll and \hyde are shown as blue and green squares, respectively, and are compared to other $z=4$ galaxies in the same field \citep{straatman2015}. Star-forming galaxies are shown in blue, quiescent galaxies in red. The half-light radii are derived from the \hubble F160W imaging (hence, rest-frame $U$ band) for all galaxies except \hyde, for which the radius is that of the dust continuum. \label{FIG:mass_size}}
\end{figure}

\hyde has very compact dust emission, $r_{1/2}<1\,\kpc$. From the absence of NIR emission in the outskirts of the galaxy (as could have been detected by \hubble), we assumed that dust is well mixed with the stars, and therefore that the stellar size must be comparable to the dust size. As shown in \rfig{FIG:mass_size}, for a stellar mass above $4\times10^{10}\,\msun$ at $3.4 < z < 4.2$, only one or two of the $14$ other star-forming galaxies (SFGs) in ZFOURGE with \hst coverage have $r_{1/2}<1\,\kpc$; instead, this size is typical for quiescent galaxies (\citealt{straatman2015}; and see also \citealt{allen2017}). The size of \hyde is in fact remarkably similar to that of \jekyll ($r_{1/2} = 0.49\pm0.12\,\kpc$; \citealt{straatman2015}).

Similarly compact SFGs have been found at higher redshifts ($z\sim4.5$), albeit with $\sfr$s larger by an order of magnitude, and were interpreted as being triggered by major mergers \citep{oteo2016}. Post-starburst galaxies at $1<z<2$ have sizes similar to quiescent galaxies \citep{almaini2017}, implying that the increase in compactness must happen within a short period of time surrounding the quenching event. In fact, the relation between stellar surface density and specific $\sfr$ of all galaxies in this redshift range suggests that SFGs become compact before they quench \citep{barro2013}. However, the converse appears to be true at $z=0$: among galaxies with strong H$\delta$ absorption, only those post-starburst galaxies with no detectable star-formation have the size and morphology of quiescent galaxies \citep{wilkinson2017}. This suggests that, in the present day, the increased in compactness happens after star-formation has started to decline, and therefore that different quenching mechanisms have acted throughout the history of the Universe \citep[e.g.,][]{schreiber2016}.

In the high-redshift context, a compelling evolutionary link can be drawn in which \hyde has formed rapidly through major mergers, building up its dense stellar core, and is now on its way to become a quiescent galaxy not unlike its companion \jekyll. One could even speculate that this burst of star-formation was triggered (or, indeed, terminated) by a recent interaction with \jekyll (see \rsec{SEC:past_jekyll}). This would be consistent with the scenario suggested by the SED modeling in which \hyde has just quenched or is in the process of quenching.

Assuming that all galaxies must grow a compact core before they quench, \cite{barro2013} found that compact $z\sim2$ SFGs had to become quiescent rapidly, within $300$ to $1000\,\Myr$. \cite{straatman2015} also showed that compact SFGs remained rare both at $z\sim4$ and $z\sim3$ ($\sim7\%$ of the SFGs), supporting the idea that compact SFGs cannot remain star-forming for very long. We therefore concluded that \hyde must be observed in this brief phase leading to quenching. Since the growth of a compact core supposedly precedes quenching, this argument does not allow us to determine if \hyde has already quenched or not, only that it must be observed within $300$ to $1000\,\Myr$ of its quenching. Given that \hyde is located at least a factor two below the main sequence, it is however probable that the quenching process has already started, and perhaps even ended.

\subsection{The cause of the low $\lcplus/\lfir$ ratio \label{SEC:cplus_deficit}}

\begin{figure*}
\begin{center}
\includegraphics[width=0.49\textwidth]{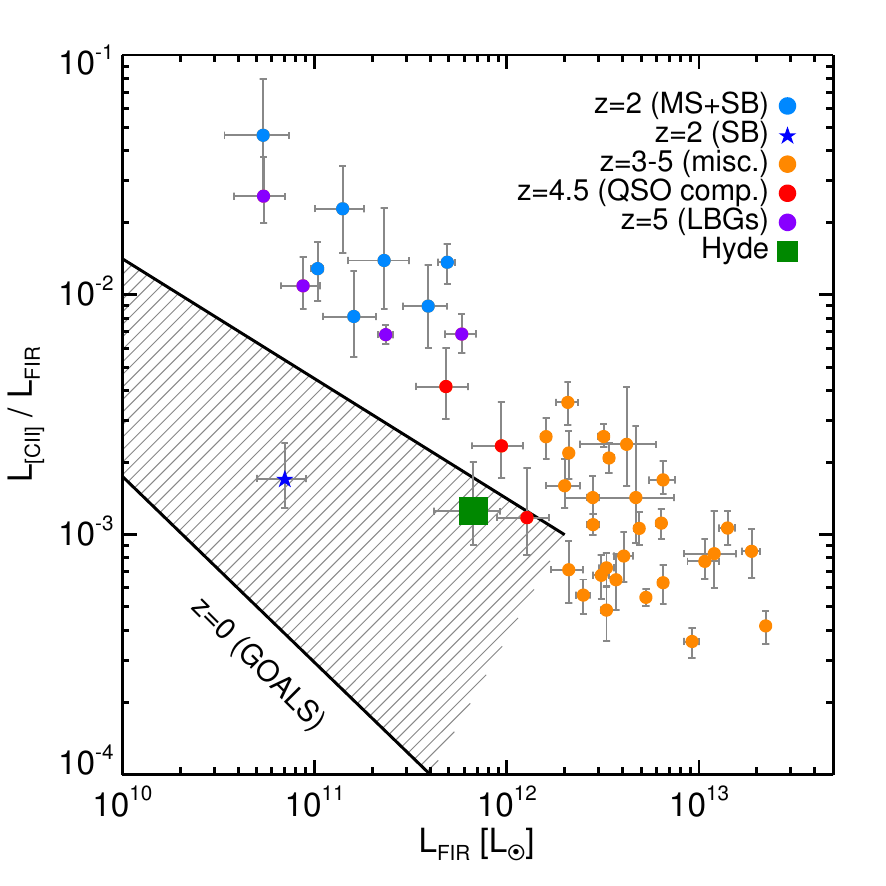}
\includegraphics[width=0.49\textwidth]{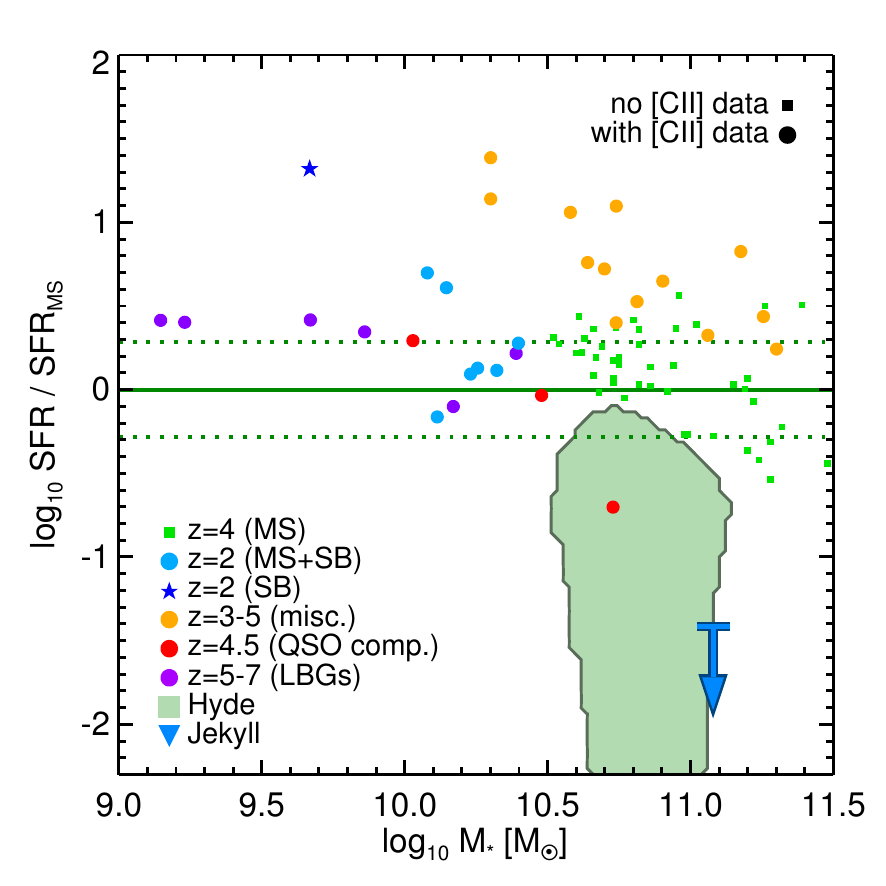}
\end{center}
\caption{{\bf Left:} Relation between the $\lcplus/\lfir$ ratio and $\lfir$. The values observed for \hyde are shown with a green square. The range of values found in $z=0$ luminous infrared galaxies are indicated with the hashed region \citep{diaz-santos2013}. High-redshift galaxies from the literature are shown with colored circles: light blue for the $z=2$ galaxies of \cite{brisbin2015}, dark blue for the lensed $z=2$ galaxy studied in \cite{schaerer2015}, red for the $z=4.5$ SMGs found near quasars in \cite{trakhtenbrot2017}, purple for the $z=5$ LBGs of \cite{capak2015}, and finally orange a collection of galaxies at $z=3$ to $5$ \citep{cox2011,debreuck2011,valtchanov2011,swinbank2012,walter2012,wagg2012,riechers2013,riechers2014,gullberg2015,oteo2016}. Galaxies from Brisbin et al.~and Gullberg et al.~with unknown magnification factors were assumed to have $\mu=10$ (the average of the published magnifications from both samples). When needed, we assumed $\lfir=\lir/1.5$. {\bf Right:} Relation between the offset from the main sequence ($\sfr/\sfr_{\rm MS}$) and the stellar mass for the galaxies on the left with measured masses. If no stellar mass estimate was available, we inferred it from the dynamical mass assuming a gas fraction of $50\%$. The value of $\sfr_{\rm MS}$ was taken from \cite{schreiber2015} at $z<3.5$ and \cite{schreiber2017-b} at $3.5\le z \le 4.5$; values at higher redshifts were estimated assuming a redshift dependence of $(1+z)^{1.5}$. On this plot, we also show with purple circles the two $z=6.6$ LBGs of \cite{smit2017}, which are detected in \cplus but not in the FIR continuum, and thus for which the $\sfr$ is based only on the UV luminosity. We also show the position of $z=4$ main sequence galaxies from \cite{schreiber2017-b} as small green squares; these galaxies have no \cplus measurement. The 90\% confidence region for \hyde is shown in light green, and the most conservative upper limit of \jekyll ($\sfr<13\,\msun/\yr$ at $3\sigma$, as obtained from $\sfr_{\rm IR}$) is shown with a blue arrow for reference. \label{FIG:cplusfir}}
\end{figure*}

We now turn to interpreting the low $\lcplus/\lfir$ ratio of \hyde in light of the above discussion.

\subsubsection{Comparison to known galaxies and trends \label{SEC:cplus_compare}}

We show in \rfig{FIG:cplusfir} (left) a compilation of $\lcplus/\lfir$ measurements for both low- and high-redshift galaxies. At the luminosity of \hyde ($\lfir \sim 10^{12}\,\lsun$), \cplus deficits are typical in the local Universe \citep[e.g.,][]{diaz-santos2013} but become rare at high redshifts \citep[e.g.,][]{brisbin2015,gullberg2015,capak2015,schaerer2015,smit2017}, except perhaps in quasars \citep[e.g.,][]{venemans2016}. In fact, searching the literature we found only three high-redshift galaxies with similar $\lcplus/\lfir$ and $\lfir \lesssim 10^{12}\,\lsun$: the lensed galaxy observed by \cite{schaerer2015} at $z\sim2$, and the galaxies found by \cite{trakhtenbrot2017} in the vicinity of $z\sim4.5$ quasars.

The galaxy studied by Schaerer et al.~is located a factor $\sim$20 above the main sequence (\citealt{sklias2014}, see also \rfig{FIG:cplusfir}, right) and is thus an extreme starburst. Such galaxies are known to have systematically lower $\lcplus/\lfir$ ratios in the local Universe \citep{diaz-santos2013}: indeed, essentially all the galaxies with $\lcplus/\lfir < 10^{-3}$ in the sample of D\'iaz-Santos et al.~have $\sfr/\sfr_{\rm MS} > 3$, thus in the local Universe \cplus deficits mainly correspond to unusually high star-formation activity. As shown on \rfig{FIG:cplusfir}, a similar trend can be observed at higher redshifts: apart from the galaxies from Trahktenbrot et al.~(which we discuss below), all galaxies with $\lcplus/\lfir < 5\times10^{-3}$ are at least a factor of two above the main sequence. In contrast, the more normal $z\sim5$ LBGs \citep{capak2015,smit2017} and $z\sim2$ galaxies \citep{brisbin2015} have no deficit. \hyde is located below the main sequence by a factor of two or more (see \rfig{FIG:cplusfir}, right), and is thus not a starburst galaxy. This suggests the cause for the deficit in \hyde must be different from that of the Schaerer et al.~galaxy and of the other luminous starbursts.

The $z=4.5$ galaxies of Trakhtenbrot et al.~are more moderate systems, as they lie within the scatter of the main sequence or even below it. These should thus be more directly comparable to \hyde, although we caution that their masses, $\sfr$ and $\lfir$ are only poorly constrained ($\lfir$ are estimated from a single flux measurement, their dust temperatures being unknown, and stellar masses are based on dynamical masses). The fact that these galaxies are satellites of bright quasars, which are believed to soon turn into quiescent galaxies, is reminiscent of the proximity of \hyde and \jekyll. It is unknown whether \jekyll did go through a quasar phase while quenching, but if super-massive black holes do reside in the cores of all galaxies, then given \jekyll's high stellar mass and compact size, a past quasar phase seems difficult to avoid (see also \rsec{SEC:sfe}). One could then speculate that the depleted \cplus emission could reflect a modification of the gas properties caused either by gravitational interaction with a massive neighbor, or by the quasar's intense radiation.

More generally, it is observed in the local Universe that the $\lcplus/\lfir$ ratio is tightly correlated with the surface density of IR luminosity ($\Sigma_{\rm FIR}$; \citealt{lutz2016,diaz-santos2017}) or equivalently with the $\sfr$ density. The origin of this relation is not clear, but it seems to hold over multiple orders of magnitudes of $\Sigma_\sfr$, at both low and high redshifts \citep{smith2017}. Despite its moderate $\lir$, \hyde has an unusually small size and its IR density is relatively high: $\Sigma_{\rm FIR} = (2.3^{+1.9}_{-1.0})\times10^{11}\,\lsun/\kpc^2$. According to the $\lcplus/\lfir$ -- $\Sigma_{\rm FIR}$ relation of \cite{lutz2016}, this value should correspond to $\log_{10}(\lcplus/\lfir) = -3.2$ to $-2.9$, which is in very good agreement with our observed ratio.

In most galaxies, $\lfir$ is a direct tracer of $\sfr$ and both quantities have thus been used interchangeably to investigate the cause of the \cplus deficit. Yet there is a physical difference between $\lfir$ and $\sfr$: $\lfir$ may only represent the obscured $\sfr$, and it may also include contributions from energy sources other than recent star-formation (e.g., older stars, or AGNs). Observationally, it is actually unknown which of $\lfir$ or $\sfr$ is best related to the \cplus luminosity.

\begin{figure}
\begin{center}
\includegraphics[width=0.5\textwidth]{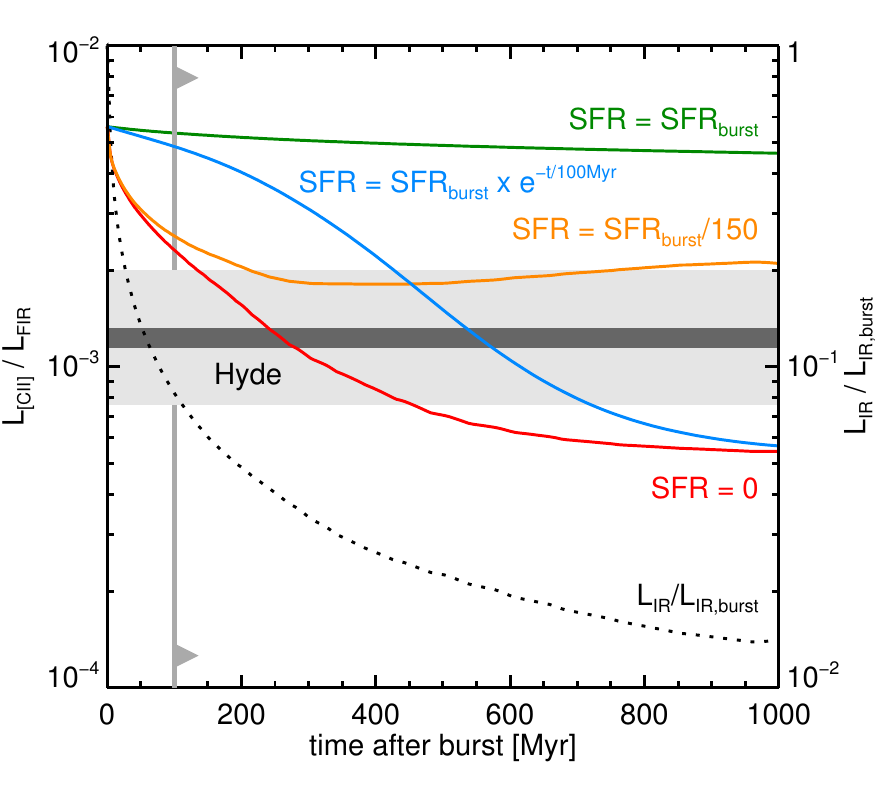}
\end{center}
\caption{Ratio of the luminosity of the \cplus line ($\lcplus$) to the far-IR luminosity ($\lfir$, $42.5$ to $122.5\,\um$) produced by a composite stellar population, as predicted by the toy model described in the text. These values correspond to a stellar population created in a single burst from $t=-100$ to $t=0\,\Myr$, and with varying amounts of residual star-formation after the burst. The red line shows the case where there is no residual star-formation, while the orange curve corresponds to a residual $\sfr$ $150$ times lower than the $\sfr$ during the burst. The blue line corresponds to an exponentially declining $\sfr$ after the burst, with timescale $100\,\Myr$. The green line shows the case of constant star-formation for reference (i.e., no end to the burst). The \cplus deficit of \hyde and its error bar are indicated with a dark gray band and shaded region in the background. Lastly, the relative decrease of $\lir$ with time in the case of no residual star-formation is shown with a dotted line. \label{FIG:c2drop}}
\end{figure}

\subsubsection{A softer radiation field?}

Interpreting the \cplus line flux from the point of view of physical conditions in the interstellar medium (ISM) is not straightforward since \cplus can originate from multiple phases of the ISM \citep[e.g.,][]{stacey1991,madden1993}, and there are therefore a number of ways to explain a deficit \citep{malhotra2001}. In nearby galaxies, the majority of the \cplus emission originates from photo-dissociation regions (PDRs; \citealt{stacey2010}). In this environment, carbon is easily ionized, and the excitation of the fine-structure \cplus emission is provided by collisions with gas particles, themselves heated by interactions with free electrons \citep{tielens1985,stacey2010}. These electrons are extracted from small dust grains and polycyclic aromatic hydrocarbon (PAH) molecules through the photoelectric effect of far UV photons emitted by nearby young stars \citep{weingartner2001}. Therefore the \cplus emission is ultimately tied to the ambient far-UV radiation.

One possible explanation for a low $\lcplus/\lfir$ ratio would thus be a softer stellar radiation field (\citealt{nakagawa1995,luhman1998,kapala2017,lapham2017}): \cplus is depressed by a lack of PAH-ionizing photons ($E > 6\,{\rm eV}$, mostly emitted by O and B stars), while residual IR emission is still produced by heating from less energetic photons (mostly produced by intermediate-age stars). This would imply a recent diminution of star-formation, which we quantify in the next paragraph. As noted in \cite{malhotra2001}, this scenario cannot be the right explanation for all the \cplus-deficient galaxies since it would imply that all starburst galaxies are observed after an substantial reduction of their star-formation activity. But in our case it is consistent with \hyde being located below the main sequence, as well as with the model of \cite{diaz-santos2017}: at $\tdust=31\,\kelvin$ and $\log_{10}(\lcplus/\lfir) = -2.91$, they predict that only half of the $\cplus$ emission is coming from dense PDRs; the rest of the emission is produced in regions of hot, diffuse ionized gas (see also \citealt{croxall2017}). There, ionization can be caused by sources other than young stars, such as post-AGB stars, shocks, or an active galactic nucleus (see \citealt{annibali2010} and references therein).

\subsubsection{A toy model using composite stellar populations}

We thus built a toy model to estimate the timescale over which a reduced $\sfr$ could cause an observable \cplus deficit (see also \citealt{kapala2017} where a similar approach was used). Since  the \cplus emission in PDRs is ultimately modulated by the ionization of dust grains and PAHs, it will respond to the incident flux of photons with $6 \lesssim E < 13.6\,{\rm eV}$ ($912 < \lambda \lesssim 2000\,\AA$; \citealt{stacey2010}), so $\lcplus^{\rm PDR} \propto \lion$. Since \hyde is strongly attenuated (see \rsec{SEC:stellar_pop}), one can assume that $\lir = \lbol$, and since $\lfir \simeq 0.61\,\lir$ for \hyde, the ratio $\lion/\lbol$ is a proxy for $\lcplus^{\rm PDR}/\lfir$.

Considering a stellar population formed in a $100\,\Myr$ burst followed by varying amounts of residual star-formation, including a continuation of the burst, we computed the time dependence of $\lion$ and $\lbol$ using the \cite{bruzual2003} stellar population synthesis model (the composite stellar population was built using FAST++). Assuming stars always remain in their birth cloud, we first estimated an empirical conversion between $\lion$ and $\lcplus^{\rm PDR}$. For a system forming stars at a constant rate, after $100\,\Myr$ the ratio $\lion/\lbol$ is equal to $0.68$ (this value decreases only mildly with time afterwards, reaching $0.59$ after $600\,\Myr$). This should correspond to the average $\lcplus^{\rm PDR}/\lfir$ ratio for star-forming galaxies. \cite{diaz-santos2017} show that, for a galaxy of $\tdust\sim30\kelvin$, the $\lcplus^{\rm PDR}/\lfir$ ratio is on average $0.9\%$, from which we derived $\lcplus^{\rm PDR} = 7.5\times10^{-3}\,\lion$. For the diffuse ionized gas component, this same model predicts $\lcplus^{\rm ion}/\lfir = 0.05\%$, which we assumed is independent of the stellar radiation field. The observed $\lcplus/\lfir$ is then the sum of these two components:
\begin{align}
\frac{\lcplus}{\lfir} = \frac{\lcplus^{\rm PDR}}{\lfir} + \frac{\lcplus^{\rm ion}}{\lfir} = 1.3\times10^{-2}\,\frac{\lion}{\lbol} + 5\times10^{-4}\,. \label{EQ:c2}
\end{align}

Based on the sample of \cite{diaz-santos2017}, we expect the scatter in both terms of this formula to be of the order of $0.2\,\dex$. We finally stress that the above equation only applies to galaxies with $\tdust\sim30\,\kelvin$, the numerical constants being dependent on the temperature.

\subsubsection{Comparison with the deficit of \hyde}

The evolution of $\lcplus/\lfir$ predicted by this simple model is shown in \rfig{FIG:c2drop} for various star-formation histories following the initial burst. In the case with no residual star-formation, the model predicts the $\lcplus/\lfir$ ratio decreases by a factor $3$ about $100\,\Myr$ after the burst. If instead star-formation continues after the burst at a rate even $150$ times lower than during the burst, the deficit barely reaches that observed in \hyde, suggesting that little on-going star-formation is allowed. If the decline in $\sfr$ is more gradual, with $\sfr \sim e^{-t/\tau}$ and $\tau=100\,\Myr$, a drop of a factor $3$ is reached $400\,\Myr$ after the burst. Therefore, deficits such as the one we observe can be explained if the galaxy is observed more than $100\,\Myr$ after the end of an intense star-formation episode, with an additional lag if star-formation was reduced gradually rather than immediately. On the other hand, these timescales could be shorter if stars were allowed to leave their birth cloud early (e.g., after $100\,\Myr$; \citealt{charlot2000}), or if the deficit of \cplus is only partly caused by a reduced star-formation (in which case the ``baseline'' value for a constant $\sfr$ would be lower to begin with, and it would take less time to reach the observed value).

This is consistent with the constraints on the quiescent time for \hyde obtained from the SED modeling, and would favor the scenario where the galaxy has just stopped forming stars. In the case of no residual star-formation, the $\lir$ observed $100\,\Myr$ after the burst is $12$ times lower than the peak value during the burst (see dotted line in \rfig{FIG:c2drop}): for \hyde, this would imply a peak $\sfr$ of $\sim1000\,\msun/\yr$ leading to a formation timescale of only $50\,\Myr$, which is also consistent with the SED modeling and observation of strong starbursts at higher redshifts \citep[e.g.,][]{daddi2009-a,riechers2013,oteo2016}. This also implies that the galaxy cannot be observed too late after a burst, in which case the residual dust continuum emission would be weak, and the inferred $\sfr$ during the burst would reach values that have never been observed ($\sfr > 10^{4}\,\msun/\yr$).

Finally, if we were to directly apply \req{EQ:c2} to the range of SFHs allowed by the SED modeling, including the expected scatter, we would predict an observed $\log_{10}(\lcplus/\lfir)$ ranging from $-2.75$ to $-1.98$ for the quenched SFHs ($t_{b<1\%} > 50\,\Myr$), which encompasses the measured value within the error bar, and $-2.43$ to $-1.85$ for the star-forming SFHs ($t_{b<30\%} = 0$), which is too high by at least $2.5\sigma$. Our toy model is fairly rudimentary so this comparison can only be indicative, but it does suggest that the star-forming solutions are disfavored.

These conclusions could be validated by observing other far-IR lines which uniquely trace diffuse ionized gas, such as $[\ion{N}{ii}]_{205}$, and determine what fraction of the \cplus is actually produced in PDRs without relying on the $z=0$ results of \cite{diaz-santos2017}.

\subsection{Star-formation efficiency \label{SEC:sfe}}

To understand the star-formation processes in \hyde and determine which process may be at play, it is useful to estimate its star-formation efficiency and gas fraction.

\subsubsection{Estimating the efficiency}

From the results of the SED modeling, we constrained the current $\sfr$ of \hyde to be less than $119\,\msun/\yr$ at 90\% confidence, and possibly zero. Given the size measured in the band 8 image, this translates into an 90\% confidence upper limit on the $\sfr$ surface density, $\Sigma_\sfr = (\sfr/2)/(\pi\,{r_{1/2,\rm dust}}^2) < 61\,\msun/\yr/\kpc^2$. If we consider the models where \hyde has quenched, this could be further reduced to $\Sigma_\sfr < 5.1\,\msun/\yr/\kpc^2$.

To compute the gas surface density, we can assume the \cplus line traces the geometry of the gas in the galaxy, but estimating the gas mass is harder. We can follow two independent approaches: first, using the dust mass and assuming a gas-to-dust ratio, and second, using the \cplus line luminosity.

The far-IR SED of \hyde does not precisely constrain the dust mass, but we can nevertheless use it to obtain an order-of-magnitude estimate of the gas mass. In \cite{schreiber2016}, we have calibrated the gas-to-dust ratio ($\delta_{\rm GDR}$) of our adopted dust library against CO and \ion{H}{i} measurements of local galaxies; these gas-to-dust ratios are higher than would be obtained with other dust models, such as the \cite{draine2007} model, since the assumed dust chemical composition is different. We found $\delta_{\rm GDR} = (155\pm23)\times(Z_\sun/Z)$, where $Z$ is the gas-phase metallicity, with a residual scatter of $0.2\,\dex$. The metallicity of \hyde is unknown, however given its large dust content it is probably close to solar. In all the following calculations we assumed solar metallicity, with an uncertainty of a factor two. Folding in all these uncertainties, the dust mass translates to a gas mass of $\mgas^{\rm dust} = 5.9^{+8.6}_{-3.6}\times10^{10}\,\msun$. This mass includes helium and hydrogen, the latter in both the molecular and atomic phases.

Alternatively, we can use the empirical \cplus--CO(1-0) correlation derived for $z=2$ galaxies in \cite{gullberg2015} and $\alpha_{\rm CO} = 2\,\msun/(\kelvin\,\kms/\pc^2)$ (from \citealt{swinbank2011}, which we assume scales as $1/Z$, see \citealt{leroy2011}). Taking into account the scatter in the \cplus-CO(1-0) relation and the metallicity, we obtained $\mgas^{\rm [\ion{C}{ii}]} = 2.9^{+3.7}_{-1.6}\times10^{10}\,\msun$. This mass should only include the molecular hydrogen, but it is nevertheless consistent with the dust-based value.

Averaging these two independent estimates with inverse variance weighting, we finally obtained a gas mass of $\mgas = 3.6^{+4.3}_{-1.9}\times10^{10}\,\msun$, which is well within the range allowed by the dynamical and stellar masses (see \rsec{SEC:disk_model}). Given the upper limit on the $\sfr$, this translates into a lower limit on the depletion time $t_{\rm dep} = \mgas/\sfr > 144\,\Myr$, or $1.8\,\Gyr$ if \hyde has quenched.

Using the \cplus spatial extent, we derived $\Sigma_{\rm gas} = (\mgas/2)/(\pi\,{r_{1/2,[\ion{C}{ii}]}}^2) = (1.0^{+1.8}_{-0.6})\times10^4\,\msun/\pc^2$. This is among the largest values ever measured: for example the $z=6.3$ ``maximum starburst'' HFLS-3 has $\Sigma_{\rm gas} = 1.4\times10^{4}\,\msun/\pc^2$ \citep{riechers2013}. Yet, this same galaxy has an $\sfr$ surface density an order of magnitude higher than \hyde: $\Sigma_\sfr = 600\,\msun/\yr/\kpc^2$. Evidently, star-formation in \hyde must be less efficient than in HFLS3.

Following \cite{daddi2010-a}, we quantified the star-formation efficiency as $\varepsilon_{\rm rot} = \Sigma_{\sfr}/(\Sigma_{\rm gas}/t_{\rm rot})$, where $t_{\rm rot}$ is the rotation period of the galaxy as derived from the \cplus line kinematics (\rsec{SEC:disk_model}): $t_{\rm rot} = 8.4^{+7.9}_{-2.8}\,\Myr$. This leads to $\Sigma_{\rm gas}/t_{\rm rot} = (1.1^{+2.3}_{-0.7})\times10^3\,\msun/\yr/\kpc^2$. Considering all models allowed by the SED fits, we found $\varepsilon_{\rm rot} < 0.13$ at $90\%$ confidence. \cite{daddi2010-a} observed a typical $\varepsilon_{\rm rot} \sim 0.5$ to $0.7$ in our range of $\Sigma_{\rm gas}/t_{\rm rot}$, and our upper limit is about a factor four lower than these values. If we only consider models where \hyde has quenched, the upper limit on $\varepsilon_{\rm rot}$ drops to $0.01$, which is more than a factor $40$ below the average.

An alternative definition of the star-formation efficiency uses the free-fall time $t_{\rm ff} = \sqrt{{r_{\rm 1/2, [\ion{C}{ii}]}}^3/(2\,G\,\mgas)} = 1.2^{+0.9}_{-0.6}\,\Myr$. This yields upper limits of $\varepsilon_{\rm ff} < 0.023$ and $0.0019$ respectively, while typical values in star-forming galaxies of similar $\Sigma_{\rm gas}/t_{\rm ff}$ are of the order of $0.01$ \citep{krumholz2012}. Therefore, with this definition the star-forming solutions are compatible with a standard efficiency, while quenched models are a factor five lower than normal.

\subsubsection{Possible interpretations}

Considering the entire parameter space allowed by the SED modeling for \hyde, it appears the star-formation efficiency is relatively low, at least a factor two lower than the normal value, and could be much lower if \hyde has recently quenched. At the same time, we have shown that the depletion time must be at least $140\,\Myr$, and possibly larger than $1\,\Gyr$. Since we have provided evidence in the previous sections that a recent quenching could be the correct interpretation of our observations, we need to understand how the galaxy could have quenched while keeping substantial reservoirs of inactive gas.

While quenched (or early-type) galaxies typically have very low gas fractions \citep[e.g.,][]{combes2007,young2014,sargent2015}, several recent studies have reported the detection of non-star-forming gas reservoirs in quenched and post-starburst galaxies, at both low and high redshifts \citep[e.g.,][]{davis2014,alatalo2014,alatalo2015,french2015,suess2017,lin2017}, so this is not a new concept. In particular, stacking $z\sim2$ quiescent galaxies on sub-millimeter images, \cite{gobat2017} found that their galaxies show a non-negligible amount of dust and gas despite their low star-formation activity. The average gas fraction they obtained ranges from $\mgas/\mstar = 0.04$ to $0.13$, which is substantially higher than the value for local early type galaxies, $\mgas/\mstar < 0.007$.

Gobat et al.~explained this low efficiency using ``morphological quenching'' \citep[see][]{martig2009}: the presence of a dense, spheroidal stellar component at the center of galaxies creates additional shear and stabilizes the gas, thus preventing star-formation. A similar interpretation was put forward in \cite{suess2017}. Given that \hyde has a stellar density similar to that of $z\sim4$ quiescent galaxies (see \rsec{SEC:size} and \rtab{TAB:props}), this would be a probable scenario. The numerical simulations of Gobat et al.~suggest morphological quenching should happen when the gas fraction decreases below $\sim$$20\%$, which takes about $2\,\Gyr$ after the main burst in their simulation. Such long timescale seems inconsistent with a recent quenching for \hyde, however it is likely to vary from one galaxy to the next. The current gas fraction of \hyde is only constrained within $\mgas/(\mgas + \mstar) = 12$ to $70\%$, so we cannot determine whether it has reached this $20\%$ threshold or not.

Alternatively, galaxies as massive as the pair we study in this paper are expected to host super-massive black holes with masses as high as $M_{\odot} \sim 10^{9}\,\msun$ \citep[e.g.,][]{reines2015}. These galaxies should have shone as bright quasars during the period in which their central black holes grew at a rapid rate \citep[e.g.,][]{trakhtenbrot2011}, and the resulting radiation is believed to trigger powerful winds which can effectively quench star-formation \citep{silk1998,king2003,cattaneo2009}. If \jekyll and \hyde do host such black holes, they were not active the moment they were observed, but it is possible that \jekyll has experienced such a phase when it quenched, $200$ to $650\,\Myr$ prior to observation.

One could also speculate that the quenching (or reduced star-formation) of \hyde has been caused by the feedback from its own super-massive black hole, which had stopped accreting at the time of observation; the quasar may in fact have been triggered by tidal interaction with \jekyll. Feedback from a central black-hole is mainly thought to act by driving powerful winds, expelling gas out of the galaxy and thus preventing it from forming stars \citep[e.g.,][]{silk1998,king2003,faucher-giguere2012,costa2014}. This scenario would not match the relatively large gas reservoir observed in \hyde. However, recent simulations of quasars implementing radiative transfer show that star-formation in compact high-redshift galaxies can be suppressed by radiation pressure on the dust, without removing the gas (Costa et al.~in prep.). In this picture, the gas is instead redistributed in the galaxy: radiation pressure erases local over-densities, thus reducing the star formation rate, without substantially altering the gas mass or the size of the galaxy. This hypothesis could be investigated further with a more robust mapping of the gas distribution in the galaxy, for example with high-resolution CO or [\ion{C}{i}] imaging. Deeper \cplus observations could also probe the large-scale environment and reveal whether or not gas has been expelled out of the galaxy \citep[e.g.,][]{cicone2015}.

In an alternative scenario, the bright quasar in \jekyll inhibited star-formation not only in its host galaxy, but also in \hyde, e.g., through radiation pressure on dust \citep{ishibashi2016}. Linearly extrapolating the line-of-sight velocity of \hyde back in time to the point where \jekyll has quenched suggests both galaxies were then separated by a projected distance of $100$ to $300\,\kpc$, such that this scenario appears implausible. However, given its current separation of $\sim3\,\kpc$, it is conceivable that a recent short quasar even in \jekyll indeed reduced the $\sfr$ in \hyde.

Finally, we note that the cessation of star-formation in \hyde (if any) may not be final. Given the proximity with \jekyll, it is possible that these galaxies have already interacted in a recent past. In this case, we could be observing \hyde in a temporary episode of quiescence, a few hundred $\Myr$ after an efficient burst. While the star-formation efficiency can fall below the normal value in this instance \cite[e.g.,][]{fensch2017}, the gas density must also decrease and it is difficult to imagine it being even larger than we currently observe.

Drawing firmer conclusions would require a better measurement of the gas mass, for example using the CO(1-0) or [\ion{C}{i}] line luminosities \citep[e.g.,][]{bothwell2017}. A more direct measurement of the $\sfr$, for example with H$\alpha$ or the Paschen series (which will be reachable with the {\it James Webb} Space Telescope) or high-J CO lines, would also settle the question of whether \hyde has truly quenched or not.

\subsection{\hyde as a probe of \jekyll's past \label{SEC:past_jekyll}}

The constraints from our SED modeling show that \hyde has formed over a timescale comparable to that of \jekyll, but has done so at a later stage. Both galaxies are otherwise surprisingly similar: they have comparable stellar masses and sizes, and have evolved in the same environment. It is therefore tempting to regard \hyde as a good representation of what \jekyll has been shortly after (or before) having quenched. This idea is supported by the fact, demonstrated in \cite{straatman2014}, that the star-formation in the progenitors of the $z\sim4$ quiescent galaxies must be strongly obscured, because the space density of UV-bright galaxies with the required $\sfr$s are too low by orders of magnitude.

In this context, any conclusion we can draw on the state of \hyde can be translated to the progenitor of \jekyll, making this system a unique laboratory to study the process of quenching.

For example, if this hypothesis is true, the fact that \hyde may have quenched while still harboring substantial reservoirs of gas implies that \jekyll should also contain some amount of gas. The fraction of this gas in the molecular phase may be low, given that little on-going star-formation is presently allowed, therefore it could prove challenging to detect in CO. Tracers of atomic gas, such as [\ion{C}{i}] or the dust continuum, may be more adequate. Currently, because the dust temperature of \jekyll is unknown, constraints based on the dust mass are loose. Conservatively assuming the low average temperature found by \cite{gobat2017} in $z\sim2$ quiescent galaxies and using the same gas-to-dust ratio as adopted in \rsec{SEC:sfe} for \hyde, the non-detection of \jekyll in the band 8 image translates into $\mgas < 3.5\times10^{10}\,\msun$ ($90\%$ confidence), or a gas fraction less than $25\%$. This is within the range of gas fractions allowed for \hyde.

We finally emphasize that both galaxies were found in a cosmological survey of small area ($363\,{\rm arcmin}^{-2}$). Consequently, while \jekyll is among the most massive quiescent galaxies at $z\sim4$ in this survey and may thus not be representative of the quiescent population at lower masses (see \rfig{FIG:mass_size}), it cannot be an extremely rare object either (the space density of such massive quiescent galaxies is $\sim 2\times10^{-5}\,\Mpc^{-3}$; \citealt{straatman2014}). Obtaining a better understanding this pair of galaxies thus has immediate consequences for our knowledge of quenching in general.

\section{Conclusions \label{SEC:conclusion}}

We have obtained new and deep ALMA data towards the most distant known quiescent galaxy at $z\sim4$ to investigate the origin of the sub-millimeter emission detected close to the line of sight. The emission was found to originate from a separate, compact source located $0.40\pm0.008\arcsec$ away from the quiescent galaxy on the continuum image, and spectroscopically confirmed to lie at the same redshift using the \cplus line. The line was found $549\pm60\,\kms$ offset from the quiescent galaxy and displays a velocity profile of a rotating disk, demonstrating it forms a separate galaxy. Careful deblending of the \spitzer IRAC images confirmed the presence of an additional source of near-IR emission, but no counterpart was found on the \hubble images, suggesting the galaxy is strongly obscured. We dubbed the quiescent and dusty galaxies \jekyll and \hyde, respectively.

Modeling of the sub-millimeter emission showed that \hyde has a moderate infrared luminosity, corresponding to an obscured $\sfr_{\rm IR}\sim100\,\msun/\yr$. Full modeling of the UV-to-FIR emission confirmed extreme levels of obscuration of the stellar light, with $A_{\rm V}\sim3.5$, and a stellar mass comparable to its quiescent neighbor. This modeling further revealed that the observed dust luminosity can be fully powered by intermediate-age stars, so the current $\sfr$ may be zero, but we could not exclude that the galaxy is still forming stars on the lower envelope of the main sequence. A similar analysis of \jekyll confirmed its initial characterization in \cite{glazebrook2017}, its stellar mass having decreased by only $30\%$, and the non-detection by ALMA confirms the absence of obscured star-formation.

Fitting the kinematics of the \cplus emission of \hyde with a rotating disk model yielded a fast rotation period of $\sim$$10\,\Myr$, a high rotation speed of $\sim$$700\,\kms$, and a compact size consistent with that of the dust continuum: $r_{1/2} = 0.67\pm0.14\pm\,\kpc$. This size was found very similar to that of \jekyll ($r_{1/2}=0.49\pm0.12\,\kpc$) and other quiescent galaxies at the same mass and redshift, suggesting \hyde may well be on its way to become quiescent itself.

The ratio of \cplus-to-far-IR emission in \hyde was found lower than any high-redshift galaxy of this luminosity. We created a toy model to determine the timescale on which the \cplus/FIR ratio can decrease following a cessation or reduction of $\sfr$, and found that the observed ratio could be reached about $100\,\Myr$ after the end of a burst, consistent with the hypothesis that \hyde may have quenched.

Using various estimates of the gas mass, we showed that \hyde has among the highest gas surface density observed in a galaxy, rivaling that of extreme starbursts at the same redshifts. Combined with its moderate-to-low $\sfr$ and fast rotation, this implies a particularly low star-formation efficiency. Consequently, whatever phenomenon is responsible for its low star-formation activity acts more by pressurizing or stabilizing the gas, rather than depleting the reservoirs.

We finally argue that, owing to their striking similarity of compactness, environment and, perhaps, star-formation history (only shifted by $\sim$$400\,\Myr$), \jekyll and \hyde can be viewed as two stages of quenching, and thus provide us with a unique laboratory to explore the physics of this poorly understood phenomenon.

A further understanding of this system could be achieved by obtaining a more direct measurement of the $\sfr$ of \hyde, for example using the {\it James Webb} Space Telescope, and constraining the gas mass of both galaxies in the three main phases: ionized (using $[\ion{N}{ii}]_{205}$), atomic (using $[\ion{C}{i}]$) and molecular (using CO). Lastly, high-resolution imaging of the stellar (with \jwst) and dust continuum (with ALMA) could reveal traces of interaction, determining if merging played an important role in shaping these galaxies.

\begin{acknowledgements}


The authors want to thank Alice Shapley for useful comments on the draft of this paper.

Most of the numerical analysis conducted in this work has been performed using {\tt phy++}, a free and open source C++ library for fast and robust numerical astrophysics (\hlink{cschreib.github.io/phypp/}).

This work is based on observations taken by the CANDELS Multi-Cycle Treasury Program with the NASA/ESA \hst, which is operated by the Association of Universities for Research in Astronomy, Inc., under NASA contract NAS5-26555.

This paper makes use of the following ALMA data: ADS/JAO.ALMA\#2015.A.00026.S and ADS/JAO.ALMA\#2013.1.01292.S. ALMA is a partnership of ESO (representing its member states), NSF (USA) and NINS (Japan), together with NRC (Canada) and NSC and ASIAA (Taiwan) and KASI (Republic of Korea), in cooperation with the Republic of Chile. The Joint ALMA Observatory is operated by ESO, AUI/NRAO and NAOJ.

GGK acknowledges the support of the Australian Research Council through the award of a Future Fellowship (FT140100933).
\end{acknowledgements}

\bibliographystyle{aa}
\bibliography{../bbib/full}

\appendix

\section{Simulated star-formation histories \label{APP:sfh}}

\begin{figure*}
\includegraphics[width=\textwidth]{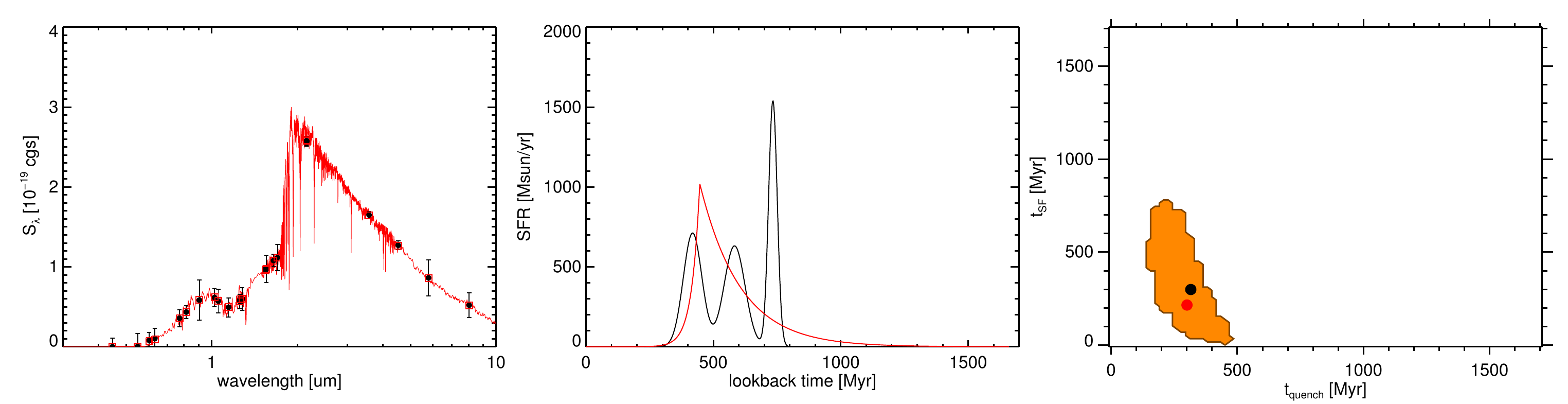}
\includegraphics[width=\textwidth]{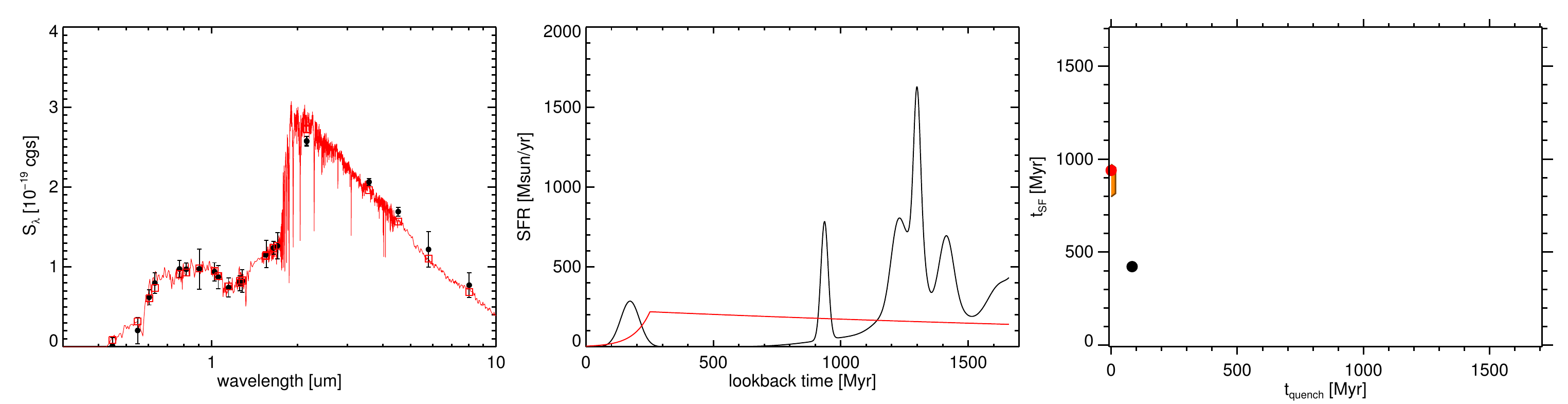}
\includegraphics[width=\textwidth]{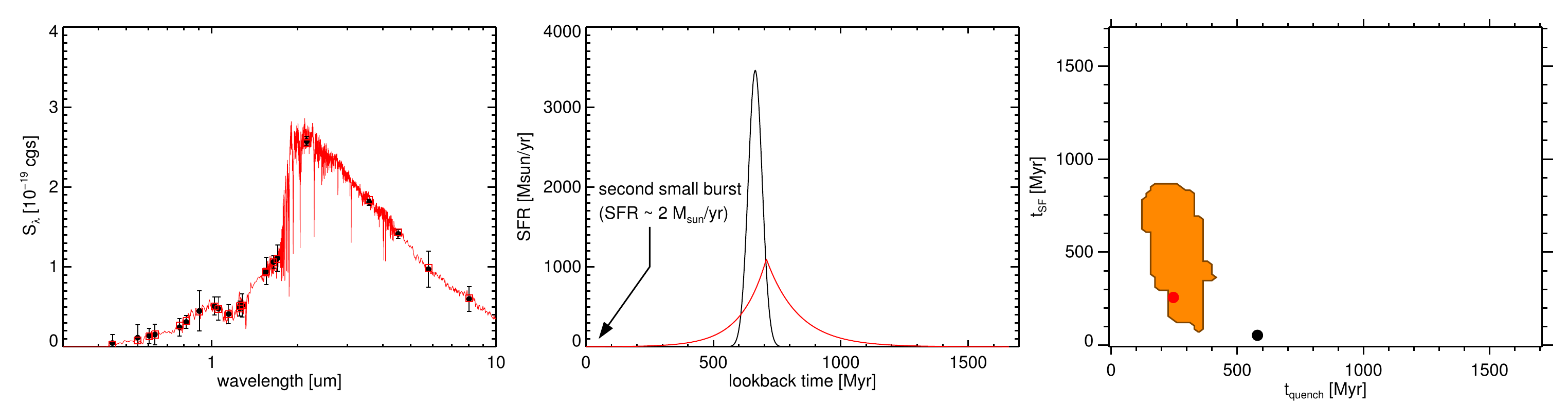}
\caption{Simulated star-formation histories for comparison with \jekyll. The first row shows a quenched star-formation history composed of three bursts. The second row shows an SFH where most of the mass was formed very early, and a small burst happened $200\,\Myr$ before observation, creating the post-starburst features. The third row shows an SFH with two burst, the last one happening when the galaxy is observed but has a very small $\sfr$. {\bf Left:} synthetic photometry for the complex SFH (black points) and best-fit model from FAST++ (red line and red squares). {\bf Middle: } complex star-formation history (black) and best-fit model (red). {\bf Right: } constraints on the formation and quenching timescales ($t_{68\%}$ and $t_{b<1\%}$) given by the model (orange area), best-fit value (red circle) and true value (black circle). \label{FIG:sim_jekyll}}
\end{figure*}

\begin{figure*}
\includegraphics[width=\textwidth]{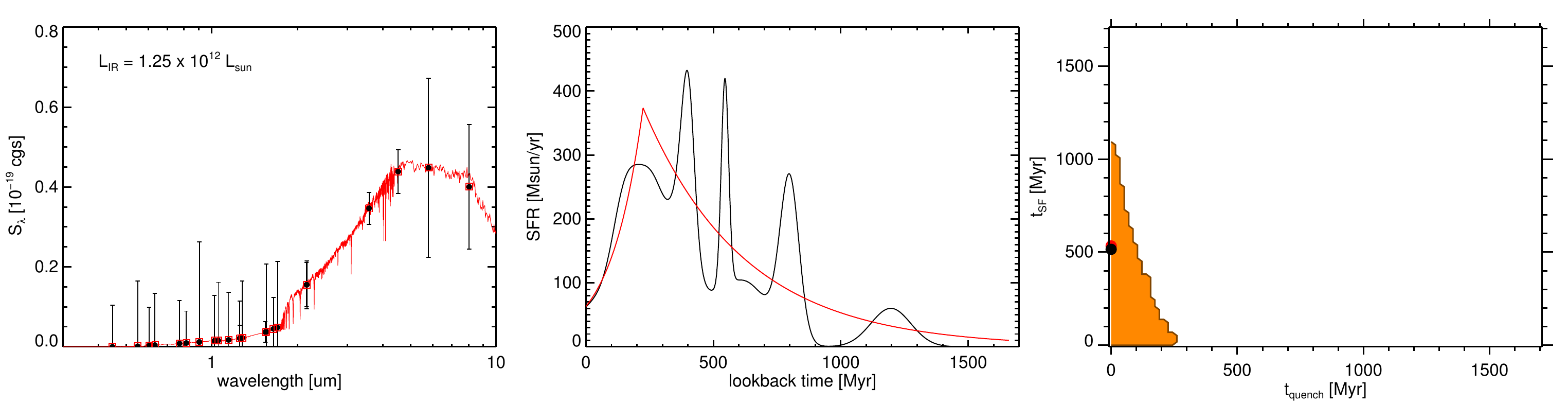}
\includegraphics[width=\textwidth]{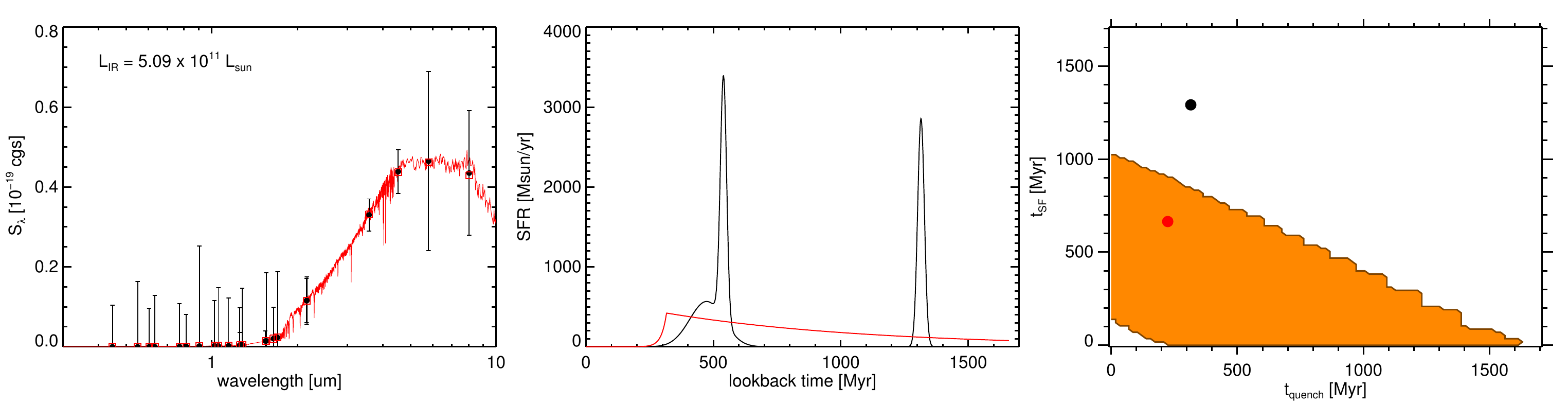}
\includegraphics[width=\textwidth]{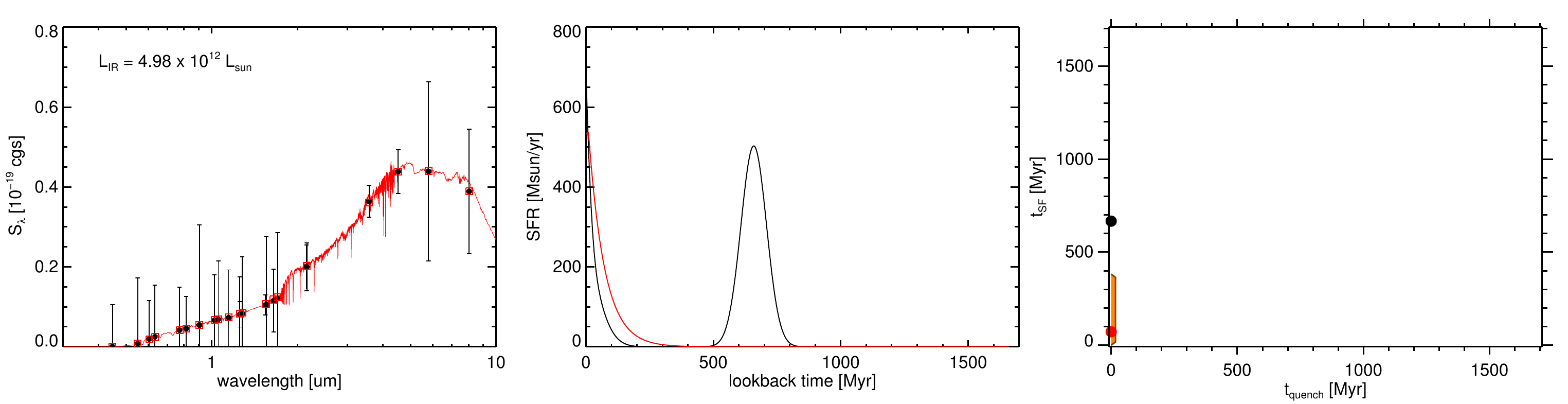}
\caption{Same as \rfig{FIG:sim_hyde}, but for comparison with \hyde. On the left column we also give the model's $\lir$ (since the SEDs are otherwise very similar). The first row shows a star-formation history which lasts steadily for several hundred million years but is composed of multiple bursts and dips, then slowly fades out $200\,\Myr$ before the observation. The second row shows an SFH with essentially two bursts happening at least $500\,\Myr$ prior observation. The third row shows an SFH with two burst, the last one happening when the galaxy is observed. \label{FIG:sim_hyde}}
\end{figure*}

The family of star-formation histories generated by \req{EQ:sfh} allows for a wide variety of scenarios, but is nevertheless simplistic. In reality, star-formation histories could be less smooth, and composed of multiple bursts. To explore whether our model still provides meaningful results in these scenario, we have generated a suite of $400$ simulated galaxies with more complex SFHs. These SFHs were created as the sum of $N$ bursts of variable peak intensity ($\sfr_b$) and duration ($\tau_b$), each burst being arbitrarily modeled as a Gaussian. The motivation for the latter is to avoid, on purpose, a functional form too similar to that assumed by \req{EQ:sfh}, in order to test how our model behaves when the true SFH has no perfect match in the grid. The time at which each burst happened ($t_b$) was drawn from a Gaussian distribution centered on a ``main burst'' time ($t_{\rm main}$) and with a given ``main burst length'' width ($\tau_{\rm main}$); $t_{\rm main}$ itself was drawn uniformly between the Big Bang and the epoch of observation, and $\tau_{\rm main}$ was drawn uniformly between $10$ and $500\,\Myr$. The number of bursts was chosen randomly from $N=1$ to $100\times(\tau_{\rm main} - 10)/490$ (with uniform probability in $\log N$) and the length of each burst was chosen randomly from $\tau_b=10$ to $200\,\Myr$ (with uniform probability in $\log \tau_b$). These values were chosen so as to provide a full coverage of the $t_{b<1\%}$ -- $t_{68\%}$ plane (i.e., \rfig{FIG:tqtsf}).

We then used the \cite{bruzual2003} stellar population models to create the corresponding SED for each simulated SFH, and generated two photometric catalogs based on these SEDs: one with $A_{\rm V} = 0.5\,{\rm mag}$, as observed for \jekyll, and another with $3.5\,{\rm mag}$, as observed for \hyde. In the first catalog the SEDs were normalized to match \jekyll's \Ks-band flux, while in the second catalog the SEDs were normalized to match \hyde's $4.5\,\um$ flux. The flux uncertainties were chosen to be the same as observed for \jekyll and \hyde, respectively.

We then ran FAST++ on these simulated SEDs with the same setup as for \jekyll and \hyde, i.e., using the model SFH of \req{EQ:sfh} (the only difference being that the simulations did not include the MOSFIRE spectrum, for simplicity). We then defined the region of the $t_{b<1\%}$ -- $t_{68\%}$ plane allowed within $\Delta \chi^2 < 2.71$ and determined, for each simulated galaxy, if the true value of $t_{b<1\%}$ and $t_{68\%}$ actually fell inside this region.

We found that $90\%$ and $95\%$ of the simulated SFHs were correctly recovered for $A_{\rm V} = 0.5$ and $3.5\,{\rm mag}$, respectively, which shows that our simplistic model is indeed able to account for more complex SFHs than that of \req{EQ:sfh}. Examples of galaxies properly recovered by our model despite their complex SFH are shown on the first row of \rfig{FIG:sim_jekyll} and \ref{FIG:sim_hyde}.

Since the percentages above are only representative of our simulated data set and not of real star-formation histories, it is illuminating to look at the few cases where the model clearly failed. We show on the second and third rows of \rfig{FIG:sim_jekyll} and \ref{FIG:sim_hyde} selected examples which illustrate the most major deficiencies. In all these cases, the SFH can be described as being composed of two distinct bursts separated by a few hundreds $\Myr$ of quiescence.

For \jekyll-like galaxies (\rfig{FIG:sim_jekyll}), we observe two different types of problems. In the first case (second row on the figure), the true SFH is composed of two main star-formation episodes, the first being very old ($t_{\rm obs} - t > 1000\,\Myr$), and the second more recent ($200\,\Myr$). The entire SFH is poorly recovered; the model tries to account for both bursts and thus obtains twice longer formation timescales compared to the true SFH. Doing so, it even fails at recovering the fully quiescent nature of the galaxy (in terms of $t_{b<1\%}$; the estimated current $\sfr$ is still fairly low). However it is also clear that the adopted model is a poor fit to the photometry: the observed \Ks and \spitzer bands are off by more than $2\sigma$. Such cases, if they happen, would be identifiable easily.

The second case is more subtle. The true SFH in this case is also composed of two bursts, the first is moderately old ($t_{\rm obs} - t \sim 700\,\Myr$) but the second has extremely small $\sfr$ and is still on-going at the time of observation. The peak $\sfr$ of this second burst is only $2\,\msun/\yr$, compared to several thousands for the main burst. The photometry is therefore dominated by the older burst, where most of the stars formed, but one can see that the on-going star-formation also leaves a noticeable trace in the UV. The model is forced to account for the main component, but it also tries to reproduce the small residual $\sfr$ coming from the second burst by using a long $e$-folding timescale for the post-burst decline. This results in a shorter quiescent time than that of the true SFH, and implies that our quenching times could be biased toward shorter values.

For \hyde-like galaxies (\rfig{FIG:sim_hyde}), we found similarly problematic SFHs. However, because the photometry there is globally of poorer $S/N$, the most stringent constraint is actually coming from the observed $\lir$. This results in a different impact on the derived parameters. Essentially, if an older burst is present in the true SFH, it is mostly ignored by the model and the most recent burst is fairly well described. In the first case illustrated on \rfig{FIG:sim_hyde} (second row), the formation timescale is enlarged to accommodate the older burst, but the quenching time is correctly captured. In the second case, the late burst happens at the epoch of observation and thus totally dominates the $\lir$; the oldest burst is completely ignored. The recovered stellar mass is a factor two lower than its true value. This is commonly referred to as the ``outshining'' phenomenon (\citealt{papovich2001}), and implies that there may have been more star-formation happening at earlier epochs in \hyde compared to what our model suggests. We emphasize that outshining is only a problem when our model SFH cannot reproduce the true SFH. In other cases, the only impact of outshining is to enlarge error bars on model parameters (in particular on the stellar mass).

\end{document}